\documentclass[12pt]{article}
\usepackage{cite}

\def\beq{\begin{equation}}
\def\eeq#1{\label{#1}\end{equation}}
\def\barr#1{\begin{equation}\begin{array}{#1}\displaystyle}
\def\earr#1{\end{array}\label{#1}\end{equation}}
\def\si{\sigma}

\def\al{\alpha}

\def\de{\delta}

\def\ta{\tau }
\def\od#1#2{\displaystyle \frac{#1}{#2}}
\def\di#1{\\[0.#1cm]\displaystyle}
\def\r#1{(\ref{#1})}

\input{epsf}
\begin{document}
\thispagestyle{empty}
\title{\bf
Radiative effects in scattering of polarized leptons by polarized
nucleons and light nuclei
}

\author{I. Akushevich$^{a)}$\footnote{on leave
of absence from the National
Center of Particle and High Energy Physics,
220040 Minsk, Belarus}, A.Ilyichev$^{b)}$, N.Shumeiko$^{b)}$ }
\date{}
\maketitle
\vspace*{-12mm}
\begin{center}
{\small {\it $^{(a)}$ North Carolina Central University,
Durham, NC 27707, USA \\ and  \\
Jefferson Lab, Newport News, VA 23606, USA\\}}
{\small {\it $^{(b)}$ National
Center of Particle and High Energy Physics,
220040 Minsk, Belarus \\}}
\end{center}

\abstract
Recent developments in the field of radiative effects in polarized
lepton-nuclear scattering are reviewed. The processes of inclusive,
semi-inclusive,
diffractive and elastic scattering are considered. The explicit formulae
obtained within the covariant approach are discussed. FORTRAN codes
POLRAD, RADGEN, HAPRAD, DIFFRAD and MASCARAD created on the basis of the
formulae are briefly described. Applications for data analysis of the
current experiments on lepton-nuclear scattering at CERN, DESY, SLAC and
TJNAF are
illustrated by numerical results.

\newpage
\tableofcontents
\section*{Introduction}
\addcontentsline{toc}{section}{Introduction}

Starting with the late nineteen sixties the main source of information
about internal nature of nucleon is lepton nuclei scattering.
Nowadays
experimental design of modern detectors opens possibility for
large set of
experimental data during short period of time. As a result the
calculation of
background radiative effects or the procedure of radiative corrections
(RC) is a quite important thing.

While considering lepton nucleon scattering it should be noted
that the main
contribution to the observables gives subprocess with one photon
exchange graphs that is called Born process. However all sets of
experimental data
consist of not only such contribution, but of other processes
appearing
during the measurement which are called background ones because of their
small contributions to the observables. Therefore the main task of data
processing in experiments on lepton nucleon scattering is
the extraction of
the one-photon exchange contribution. It leads to the necessity to
cancel
the background events from all data sets. When it cannot be made
by
experimental methods the correspond contributions can be
calculated
in the frame of some theory.  Usually a such kind of
contribution depends
on the quantities (the structure function, asymmetries), that are
observables in the given experiment. As a result we come to
the necessity of the
iteration procedure application for the extraction of
the observables on Born
level from the full set of experimental data. This procedure is usually
called as the procedure of RC of experimental data.

There are two basic methods of calculation of model independent QED
radiative correction. First one is connected with introducing of an
artificial parameter separating the momentum phase space into soft and
hard parts. One can find a classical review introducing this
formalism in ref.\cite{MoTsai,Tsai}. However the presence of the
artificial parameter is a disadvantage of this method. For the soft
photon part the calculation is performed in such
approximation when the photon energy is considered to be small
with respect to all momenta and masses in the problem.
So this parameter should be chosen as small as possible to reduce
the region evaluated approximately. From the other side it cannot
be chosen too small because of  possible numerical instabilities in
calculating of hard-photon emission.
Calculation within the approach of Mo and
Tsai was performed only for the case of unpolarized deep inelastic
scattering.
 In the end of seventies Bardin and Shumeiko
developed an approach \cite{BSh} of extraction and cancellation of
infrared divergence without introducing this artificial
parameter.
Then this approach was applied to the
calculation of electromagnetic correction to the lepton current
for deep inelastic scattering (DIS)
\cite{ABSh, Sh2el, ABK, Soroko1, KSh1,ARST, Soroko2, AkuSor,Sh}
as well as for some
approximation and taking into account the contribution of net soft
photon emission.
The exact expressions for the lowest order QED RC to polarized DIS
of polarized leptons by polarized light nuclei including targets
of spin 1 such as deuterium are presented in \cite{ASh}.
Basing on these formulae FORTRAN code POLRAD \cite{P20} are
constructed. Besides, some others tasks such as
Monte Carlo generator RADGEN \cite {RAD} was made, RC both to diffractive
vector meson electroproduction \cite{DIFF,DIFF2} and to spin-density
matrix elements in the exclusive vector meson production
\cite{AKuzh}
by this method were solved. Recently Bardin-Shumeiko approach has
being applied to experiment on polarized elastic electron-proton
scattering at TJNAF \cite{pep1,pep2}.
The comprehensive analysis of some results
obtained by \cite{MoTsai, Tsai} and \cite{BSh}
was made in \cite{KuzhSh, RAD}.

The formulae of \cite{BSh} as well as \cite{KShT} allow
along with using the parton model to estimate the value of
electromagnetic correction to the hadronic current in unpolarized
\cite{t-79}
and polarized
\cite{ShT, ShZ}
lepton nucleon DIS.

Other important source
of contributions to systematic error of experiments with high
energy scattering particles
together with electromagnetic effects
is electroweak ones.
When the square transferred momentum is high enough the
contribution of
$Z$-boson change is compared with
electromagnetic one especially on the kinematic borders.
Therefore it is useful to have the calculation of RC within the
standard model of electroweak interactions. This task
is especially actual for experiments at collider
\cite{herau,herap}.
This calculation requires choosing
renormalization scheme and fixation the gauge. Detailed description of
on-mass scheme within t'Hooft Feynman gauge were reviewed in
\cite{BHS,Holl}.

The calculation of electroweak corrections to the lepton legs for DIS
could be found in \cite {AKP,BARD,LRC}. Electroweak corrections to the
hadronic current in quark-partonic model (QMP) can be found in
\cite{BFSh1,BFSh2,BforHERA,bom_z} and \cite{AISh}
for unpolarized and polarized DIS respectively.

The calculation of
RC within the structure function
method was suggest by Kuraev and Fadin \cite{KF} and applied for DIS
\cite{KMF}. This method was used for the calculation of Compton tensor
with heavy photon in the case of unpolarized \cite{KT1} and
longitudinal-polarized \cite{KT2} fermions. The calculation of LO and NLO
corrections to radiative tail from elastic peak in DIS can be found in
\cite{AKS}.

The next step is the calculation of higher level RC $\sim \alpha
^2$.
Notice that such kind of correction to the elastic tail was
calculated in \cite{Sh2el, AKS}. As to the continuous spectrum
the obtaining of the formulae not only in exact form (like
\cite{ASh}) but and in
ultrarelativistic approximation is a quite difficult task
(first of all because of the
problem of the integration over the photon emission phase space).
Therefore we consider the correction $\sim \alpha ^2$ to the continuous
spectrum in the LO approximation. For unpolarized particles such
correction in LO and NLO approximation was found in  \cite{KF,KurSpi}.

Another important source of radiative effects in DIS is
QCD-correction appearing from gluon emission.
However QCD corrections, in spite of
photon emission
is not extracted from the structures
functions but is directly included into their definition. As a result
the structure functions start depending not only on a scaling
variable $x$ but on the transferred momentum squared.

A lot of articles dedicated to NLO QCD corrections to polarized DIS have
been published. The calculations in most of them are performed in
the assumption that the quark mass is equal to zero (see
review \cite{QCD} and reference there). However there
are some works when the authors estimate a finite quark mass
effect
for both unpolarized \cite{NE1} and polarized structure functions
\cite{NE2,AG2,TER,TERV,TERM,NE3,BT}. The main difference of these two
approaches is the method of tending the quark mass to zero and the
procedure of integration of the squared matrix element over the phase
space of an emitted gluon. In the massless approach a fermion mass
is equal
to zero {\it before } the integration while in the massive one this
quantity goes to zero {\it after} the integration and survives only in LO
terms.

Notice that QCD and QED radiative corrections (RC) having different
origins
possess some common features on the one-loop level.
If our consideration is restricted to so-called QCD Compton
process then both corrections should be described by the
identical sets of Feynman graphs. The transition from the strong
radiative effects to the electromagnetic ones could be performed
by the following replacement:
\begin{equation}
\frac 4 3 \alpha _s \rightarrow e_q^2\alpha _{QED}.
\end{equation}

Due to these facts in \cite{QCD} the mentioned above Bardin Shumeiko
method was applied to the investigation of Compton effects within
QCD. This
calculation is actual because it allows us to estimate the influence of
the finite quark mass on observables in DIS like it was done in
\cite{NE1,NE2,AG2,TER,TERV,TERM}.

In the present poster which is basing
on
ref.\cite{RAD,Soroko1,AkuSor,ASh,P20,DIFF,DIFF2
,AKuzh,pep1,pep2,LRC,LO,QCD}
some radiative effects in
polarized lepton by polarized light nuclei scattering is
reviewed. In the first and last sections the lowest  order
corrections are calculated within Bardin Shumeiko approach as well
as the next order correction for leptonic current are estimated in the
leading level by the method described in \cite{KurSpi}.
The calculation of QCD-correction
to the hadronic current is reviewed with \cite {QCD}.

\markright{ }
\section{Corrections to the lepton current in DIS}
\markright{ }
\subsection{Inclusive physics}

Let us consider the process of polarized lepton-nucleon DIS
\begin{equation}
l(k_1,\xi )+N(p,\eta )\rightarrow l'(k_2)+X \label{sc2}
\end{equation}
where $k_1$, $p$ ($\xi $, $\eta $) are the momenta
(polarization vectors) of the initial particles, $k_2$ is the final lepton
momentum and $X$ is a final hadronic state.

\begin{figure}
\vspace{2cm}
\begin{tabular}{ccccc}
\begin{picture}(60,100)
\put(-50,-110){
\epsfxsize=7cm
\epsfysize=11cm
\epsfbox{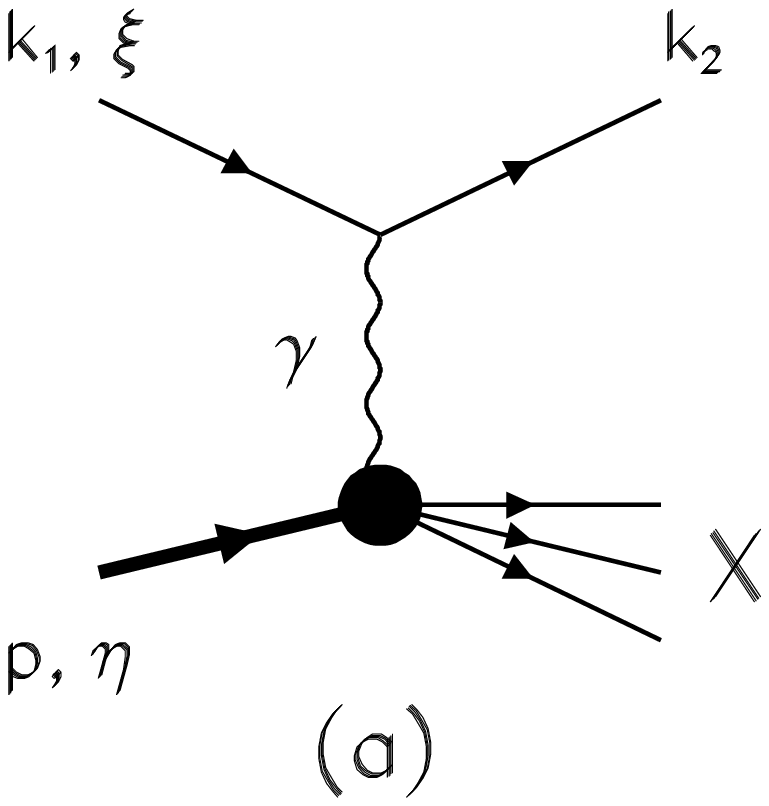} }
\end{picture}&
\begin{picture}(60,100)
\put(-25,-110){
\epsfxsize=7cm
\epsfysize=11cm
\epsfbox{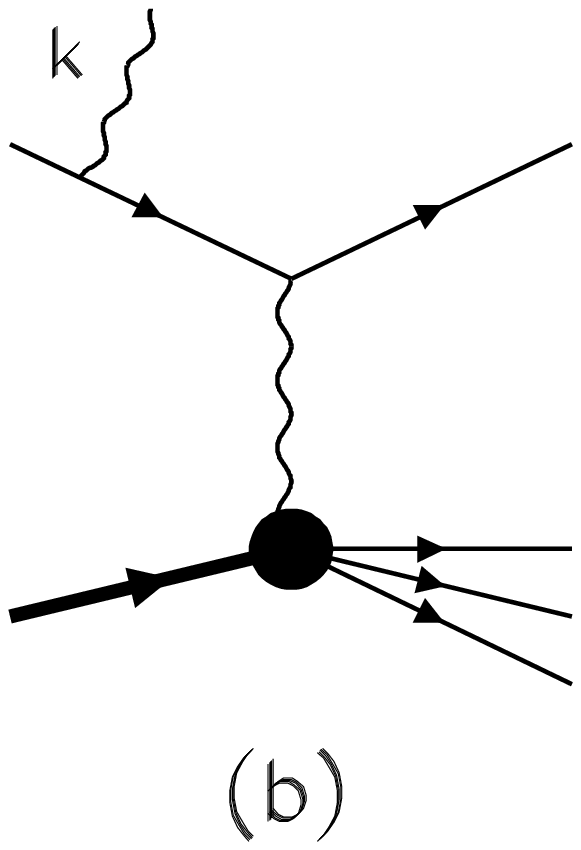} }
\end{picture}
&
\begin{picture}(60,100)
\put(-5,-110){
\epsfxsize=7cm
\epsfysize=11cm
\epsfbox{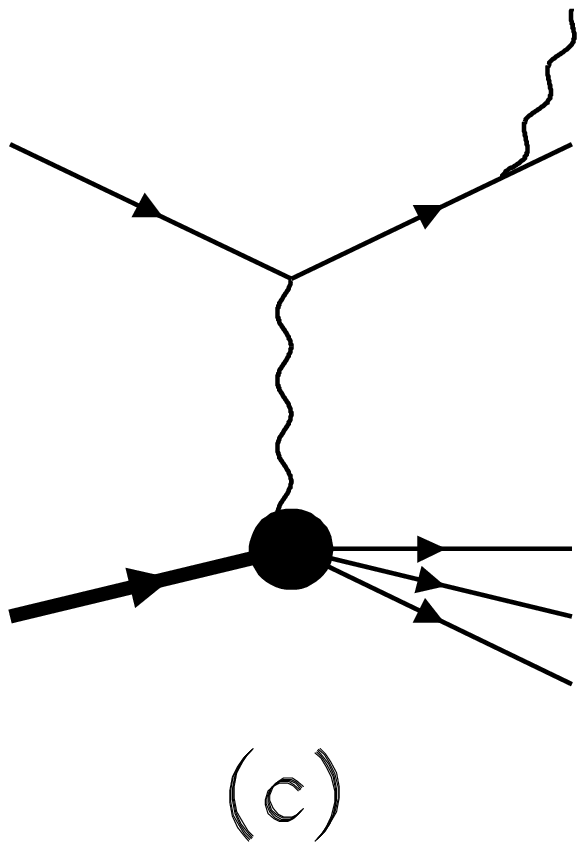} }
\end{picture}
&
\begin{picture}(60,100)
\put(15,-110){
\epsfxsize=7cm
\epsfysize=11cm
\epsfbox{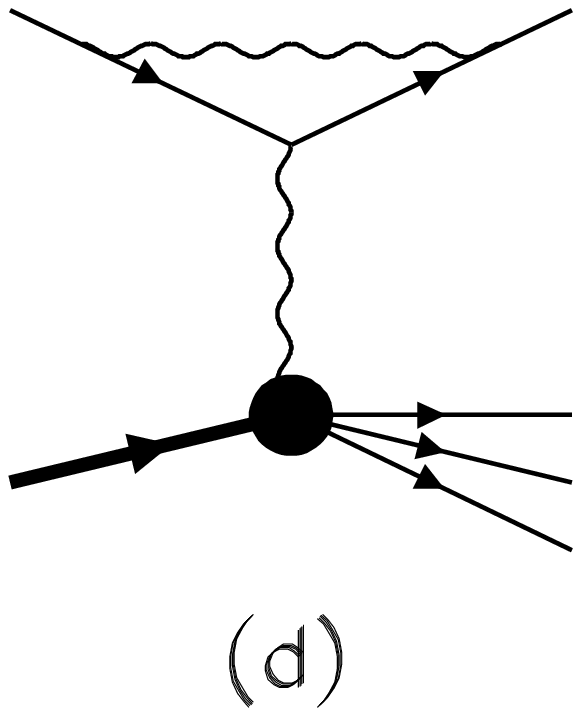} }
\end{picture}
&
\begin{picture}(60,100)
\put(35,-110){
\epsfxsize=7cm
\epsfysize=11cm
\epsfbox{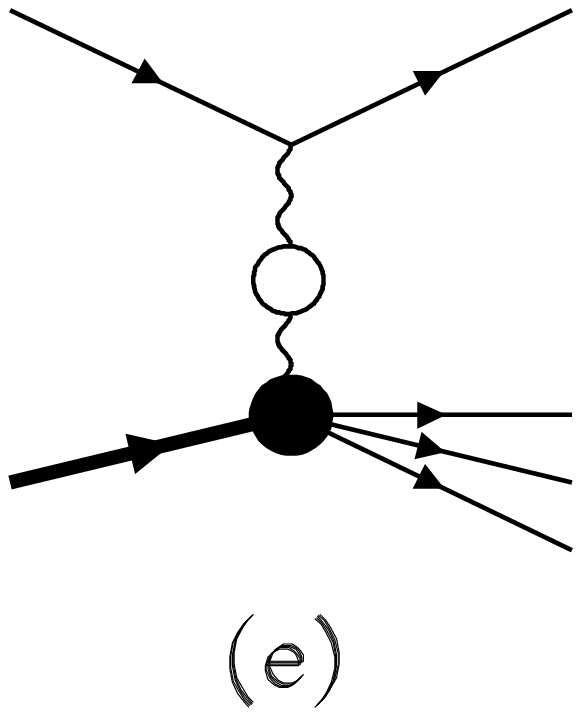} }
\end{picture}
\end{tabular}
\vspace*{-5mm}
\caption{\it
Feynman diagrams contributing to the Born and
the model independent
QED correction of the  lowest order.
The letters denote the four-momenta and polarizations of corresponding
particles.
}
\label{feyn}
\end{figure}

At the Born or one photon exchange level (fig.\ref{feyn}a) the
cross section of the considered process
can be presented by the convolution of the leptonic
tensor $L_{\mu \nu }$ having well known structure,
with the hadronic one $W_{\mu \nu }$. The latter
can be expanded into Lorentz covariants whose coefficients define
the structure function which are to be measured.

As a result the double differential Born cross section has the
form
\footnote{Now and later  $\sigma \equiv
d^2\sigma /dxdy$.}
\begin {equation}
\begin {array}{rcl}
\displaystyle
\sigma _0&=&
\displaystyle
{4\pi \alpha ^{2}\over {\lambda _{s}}}
 {SS_{x}\over Q^{4}}
 \left\{
  (Q^2-2m^{2})\Im _{1}+(SX-M^{2}Q^2){\Im _{2}\over
2M^2} \right. \\[0.5cm]
&&
\displaystyle
+ mMP_{L}
 \left(2(Q^2\;\xi \eta -q\eta \;k_{2}\xi ){\Im _{3}\over
  M^2}+ (S_{x}\;k_{2}\xi -2\;\xi p\;Q^2)
 q\eta {\Im _{4}\over M^4} \right) \\[0.5cm]
&&
\displaystyle
+  (Q^2-2m^{2})(Q^2-3(q\eta )^2){\Im _{5}\over
 M^2}+ (SX-M^2Q^2)(Q^2-3(q\eta )^{2}){\Im _{6}\over
 2M^4}     \\[0.5cm]
&& \left.
\displaystyle
 - {1\over 2}\pmatrix{Q^2+4m^2+12\;\eta k_1\;\eta
 k_2}\Im _{7} - {3\over 2}(X\;\eta k_{1}+S\;\eta k_{2})
 q\eta{\Im _{8}\over M^2}
 \right\} ,
\end {array}
\label{bt}
\end {equation}
where the kinematic invariants defined in the standard way:
\begin{equation}
\begin{array}{c}
\displaystyle
S = 2k_{1}p,\; X = 2k_{2}p = (1-y)S,
\; Q^2 =-q^2= -(k_{1}-k_{2})^{2} ={ xyS},
\\[0.5cm]\displaystyle
S_{x}= S-X,\;S_p=S+X,\; \lambda _{s}=
S^{2}-4m^{2}M^{2},
\end{array}
\label{invar}
\end{equation}
and $x=Q^2/S_x$, $y=S_x/S$ are usual scaling variables. The
explicit expression for the hadronic tensor and the generalized
structure functions $\Im_i$ can be found in Appendix A.1 of
\cite{P20}.

Now and later we deal with both longitudinal and
transversal polarized targets. At the same time the initial lepton
will be polarized longitudinally and its polarization vector can
be presented as the sum of two parts:
\begin{equation}
\xi  = {S\over m \sqrt{\lambda _{s}}}k_{1}-{2m\over \sqrt{\lambda_{s}}}p
=\xi _0+\xi _1.
\label{vecpol}
\end{equation}

Notice, that in ultrarelativistic approximation over the lepton
mass a non-zero contribution to the polarized
part of cross section (\ref{bt}) appears from $\xi _0$
as well as the second part of polarization vector $\xi _1$ gives a
non zero contribution to the cross section of DIS within
$m \rightarrow 0$ by the correction from real photon emission.

The natural approximation for high energies of the
scattering particles is ultrarelativistic one
\begin {equation}
m^2,M^2 \ll S,X,Q^2.
\end {equation}

Within this approximation the Born cross section has the form
\begin {equation}
\begin {array}{c}
\displaystyle
{\sigma _0^{||}}={4\pi \alpha ^{2}S\over Q^4}
((F_{1}-{Q_{N}\over 3}b_{1})xy^{2}+(F_{2}-{Q_{N}\over 3}b
 _{2})(1-y)
-P_{L}P_{N}xy(2-y)g_{1})
\end {array}
\label {bl}
\end {equation}
for longitudinal polarized nucleus and
\begin {equation}
\begin {array}{rl}
\displaystyle
\sigma _0^{\bot}=&
\displaystyle
{{4\pi \alpha ^{2}S\over Q^4}}
((F_{1}+{Q_{N}\over 6}b_{1}
 )xy^{2} + (F_{2}+{Q_{N}\over 6}b
 _{2})(1-y)
\\[0.5cm]&\displaystyle
-2P_{L}P_{N}{x\sqrt {xy(1-y)}M\over \sqrt {S}}
(yg_{1}+2g_{2}))
\end {array}
\label {br}
\end {equation}
for transversal polarized one.

\begin{figure}
\vspace{-2cm}
\hspace{10mm}
\unitlength 1mm
\begin{picture}(60,60)
\put(20,-20){
\epsfxsize=8cm
\epsfysize=8cm
\epsfbox{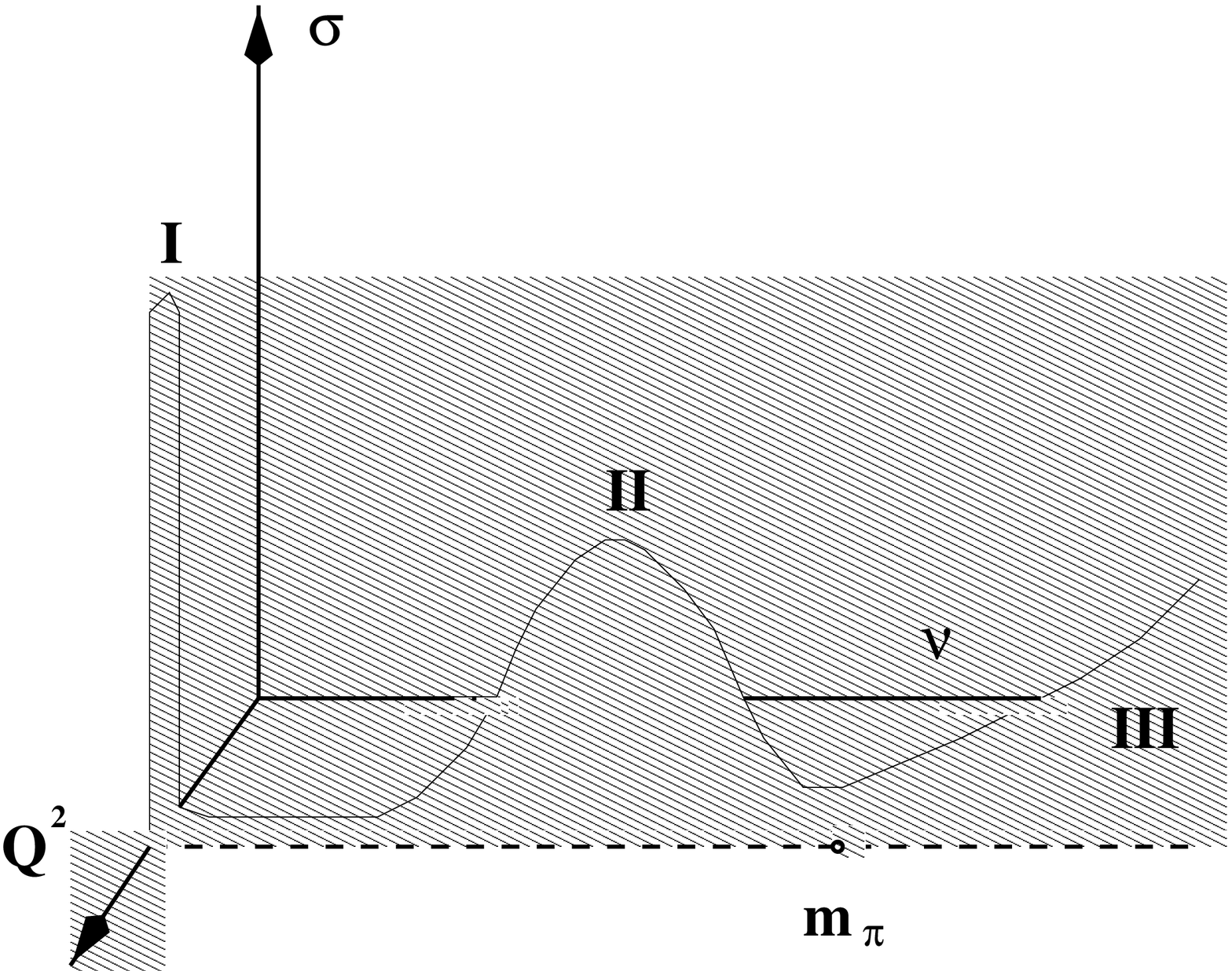}
}
\end{picture}
\vspace{20mm}
\caption{\protect\it
The double differential cross section for inclusive
lepton-nuclei scattering sketched at a certain value of $Q^2$ as
function of $\protect\nu$ (arbitrary scale). The three basic channels
are the elastic (I), the quasielastic (II), and the inelastic (III)
one.
}
\label{gr}
\end{figure}

    It is well-known that there are three scattering channels of virtual
boson ($\gamma$, $Z$) on nucleus in  dependence on the transfer
energy $\nu =E_{1}-E_{2}$,
where $E_{1}(E_{2})$  is the initial  (scattered) lepton  energy:
elastic,
quasielastic and inelastic.   Representative plot  of dependence of  the
scattering  cross  section  on  $\nu  $  and square of transfer momentum
$Q^{2}=-q^{2}$ is shown on fig.1 (only for the regions that give
sufficient contribution to RC calculation). the peak in the range
I (for $\nu = 0$, when  neglecting nuclear  recoil) corresponds
to the elastic scattering.
Then the nucleus  remains in the ground  state.  Range II
stands for the
quasielastic scattering i.e. direct collisions of leptons with  nucleons
inside nucleus.  A wide maximum in the energy spectrum originates
from the
own movement of nucleons.  Range III of inelastic scattering occurs when
the transfer energy is greater then pion threshold.

On the Born level both $\nu $ and $Q^{2}$ are fixed by the
measurements
of the scattered photon momentum. Hence, the  channel of  scattering
is  fixed too. However, on a level of RC radiated real photon momentum
is indefinite, and $\nu  $  and  $Q^{2}$ are arbitrary so each of
three channels contributes to cross section. Integration over
the photon phase  space may be presented in the plane of
$\nu  $ and $Q^{2}$. Adding the virtual photon contribution
$\sigma ^{v}$, we have for the RC cross   section
\begin {equation}
\sigma _{RC}  = \sigma ^{in}
+ \sigma ^{el}+ \sigma ^{q}+ \sigma
^{v}.
\label{eq1}
\end {equation}
Here $\sigma  ^{in}$,
$\sigma  ^{el}$, $\sigma  ^{q}$
are contributions  of radiative tails from continuous
spectrum, of the elastic scattering radiative tail,
of the radiative tail from the quasielastic  scattering
respectively.
Also the contribution of electroweak correction
calculated in the quark-parton model
is contained in
$\sigma  ^{in}$.
Both $\alpha$ and $\alpha^2$ corrections are taken into account in
$\sigma^{in,el,q,v}$. To separate the contributions we introduce
the lower index, f.e. $\sigma^{in}=\sigma^{in}_1+\sigma^{in}_2 $.

A modern approach of solving the task on RC calculation assumes  exact
calculation of the lowest order model independent correction. This
correction includes the QED processes of radiation of unobserved real
photon and one-loop diagrams to lepton line. These processes gives the
largest contribution, which
can be calculated exactly. Uncertainties of
the model independent RC can come only from fits and models used for
the structure functions.
The calculation of model dependent correction (box-type diagrams,
emission by hadrons) requires
additional assumptions about hadron interaction, so it has additional
pure theoretical uncertainties, which are hardly controlled. The
model-dependent correction is much smaller compared to leptonic radiation
because it does not include a large logarithmic term $\log(Q^2/m^2)$.
In this and last sections we concentrate on the calculation of
model independent correction as main contribution to total RC.

The explicit expression for the lowest-order QED correction
(see fig.\ref{feyn}b-e) can be found in \cite{ASh}.
Thus, for infrared free sum
$\sigma ^{in}_1$ (fig.\ref{feyn}(a,b))
and
$\sigma ^{v}_1$ (fig.\ref{feyn}(c,d))
we have
\begin {equation}
\sigma ^{v}_1 + \sigma ^{in}_1= {\alpha \over \pi }
\delta_{v}
 \sigma _{o} + \sigma ^{in}_F.
= {\alpha \over \pi }
(\delta^{IR}_{R} + \delta _{vert} + \delta _{vac}^l + \delta _{vac}^h)
 \sigma _{o} + \sigma ^{in}_F.
\label{RpV}
\end {equation}
The finite contribution from real photon emission
\begin{equation}
l(k_1,\xi )+N(p,\eta )\rightarrow l'(k_2)+\gamma(k)+X \label{sc2g}
\end{equation}
can be presented
in a following way:
\begin {equation}
\begin {array}{l}
\displaystyle
\sigma ^{in}_F = -{\alpha ^{3}y}
 \int\limits^{\tau_{max}}_{\tau_{min}}d\tau \sum^{8}_{i=1}
 \left\{ \theta _{i1}(\tau)
\int\limits^{R_{max}}_{0}{dR\over R} \right.
\left[ {\Im _{i}(R,\tau )\over (Q^2+R\tau)^2}-{\Im _{i}(0,0)\over
Q^4}\right]\\[0.5cm]
\displaystyle
\;\;\;\;\; \qquad \qquad\qquad\qquad\qquad\left.
+ \sum^{k_{i}}_{j=2} \theta _{ij}(\tau)\int\limits^{R_{max}}_{0}
dR {R^{j-2}\over (Q^2+R\tau)^{2}}\Im _{i}(R,\tau)  \right\},
\end {array}
\label{eq25}
\end {equation}
where
\begin {equation}
 R=2kp,\;\;\; \tau=kq/kp
\label{rtau}
\end{equation}
and $k$ is a momentum of emission
photon ($k^2=0$). The limits of integration are defined as
\begin {equation}
\label{rtau2}
\begin {array}{c}
 \displaystyle
R_{max} ={{W^{2}-(M+m_{\pi })^2}\over 1+\tau},  \ \
 \displaystyle
\tau _{max,min} = {S_{x} \pm \sqrt {\lambda _{Q}}\over 2M^{2}},
\\[0.5cm] \displaystyle
\lambda _{Q}= S^{2}_{x}+ 4M^{2}Q^2,\ \  W^{2}= S_{x}- Q^2 +
M^{2},
\end {array}
\end {equation}
where $m_{\pi}$ is a pion mass. The explicit expression for
functions $\theta _{ij}(\tau )$ are presented in
\cite{P20,ASh}.

The quantity $\delta ^{IR}_{R}$  appears  when the infrared
divergence is extracted in accordance with  the Bardin and
Shumeiko method \cite{BSh} from $\sigma ^{in}$.
The  virtual  photon  contribution consists of the  lepton
vertex correction $\delta _{vert}$  (fig.\ref{feyn}(d)) and  the
vacuum polarization (fig.\ref{feyn}(e)) by
leptons $\delta ^{l}_{vac}$  and  by  hadrons $\delta ^{h}_{vac}$
\cite{delvac}.
These corrections are given by formulae (20-25) of
ref.\cite{ASh}.

In the case of elastic scattering the nucleus remains in the ground
state, so for the integration variable $R$ we have an additional
relation
\begin {equation}
R=R_{el}=(S_{xA}-Q^2)/(1+\tau_A),
\label{Rel}
\end {equation}
resulting in
\begin {equation}
\sigma_1^{el}= {1\over A}{d^2\sigma ^{el}\over dx_Ady}=
- {\alpha ^{3}y\over A^2}\int\limits^{\tau_{Amax}}_{\tau_{Amin}}d\tau_A
 \sum^{8}_{i=1}\sum^{k_{i}}_{j=1} \theta _{ij}(\tau_A){
 2M^{2}_A R^{j-2}_{el}\over (1+\tau_A)(Q^2+R_{el}\tau_A)^{2}}\Im
^{el}_{i}(R_{el},\tau_A).
\label{eq23}
\end {equation}
Here invariants with the index "A"
 contain the nucleus momentum
$p_A$ instead of $p$ ($p_A^2=M_A^2$, $M_A$ is nucleus mass).
The quantities $\Im ^{el}_i$ are given in Appendix 1 of
\cite{P20}.

Quasielastic scattering corresponds to direct collisions of leptons
with nucleons
inside nucleus.  Due to
self movement of nucleons we have no additional relation like \r{Rel}.
As a result we have to integrate numerically both over $R$ and $\ta$
\begin {equation}
\sigma_1 ^{q}= - {\alpha ^{3}y\over
A}\int\limits^{\tau_{max}}_{\tau_{min}} d\tau\sum^{8}_{i=1}
 \sum^{k_{i}}_{j=1} \theta _{ij}(\tau)\int\limits^{R_{max}^q}
 _{R_{min}^q} dR
 {R^{j-2}\over (Q^2+R\tau)^2}\Im_{i}^q(R,\tau).
\label {qu}
\end {equation}

The quantities $\Im ^q_i$ can be obtained in the terms of quasielastic
structure functions (so-called response functions, see
\cite{P20} for explicit result), which have a form of the peak
for $\omega=Q^2/2M$. Due to the absence of enough experimental data the
fact is normally used for construction of the peak type approximation.
The factors at response functions are estimated at the peak, and
subsequent integration of response functions leads to results in terms
of suppression factors $S_{E,M,EM}$ (or of sum rules for
electron-nucleus scattering
\cite{LLS}):
\begin {equation}
\sigma_1 ^{q}=
- {\alpha ^{3}y\over A}\int\limits^{\tau_{max}}_{\tau_{min}}d\tau
 \sum^{4}_{i=1}\sum^{k_{i}}_{j=1} \theta _{ij}(\tau){
 2M^{2} R^{j-2}_{q}\over (1+\tau)(Q^2+R\tau)^{2}}\Im
^{q}_{i}(R_{q},\tau).
\label{eq23q}
\end {equation}

Detailed comparison of the results for quasielastic tail obtained within
different approaches can be found in \cite{ARST}.

To take into account the effect of radiation of many soft photons
a special procedure of exponentiation was applied \cite{Sh}.
It can be  realized by the following substitutions:
\barr{c}
 \si^{el}_1\rightarrow \left({y^2(1-x/A)^2\over
1-xy/A}\right)^{t_r}\si^{el}_1,
\qquad
 \si^{q}_1\rightarrow \left({y^2(1-x)^2\over
1-xy}\right)^{t_r}\si^{q}_1,
\di5
 (1+\od{\al}{\pi}\delta_v)\si_o \rightarrow
 \exp\left(\od{\al}{\pi}\de_{inf}
 \right)\bigl(1+\od{\al}{\pi}(\delta_v-\delta_{inf})\bigr)\si_o,
\earr{00721}
where $t_r= \od{\al}{\pi}(l_m-1)$.

\subsection{Iteration procedure of data processing}

In this  subsection
the procedure of radiative
correction to extract the structure function $g_1(x)$
from measured spin asymmetry
$A_{1i}^m$ is considered. The spin average structure
functions are considered to be constant and
$g_{2}(x)$  equals to  0.
The measured asymmetry is defined as
\begin {equation}
A^m_1 = {g_1\over F_1} + \Delta A_1(g_1),
\label{53}
\end {equation}
where
 the radiative correction to asymmetry
$\Delta A_1$
can be written in terms of spin-average and spin-dependent parts
($\si^{u,p}$) of cross sections (\ref{eq1})
\begin {equation}
\Delta A_1=
{\sigma_0^u\bigl(\sigma_p^{in}(g_1)+\sigma_p^{q}+\sigma_p^{el}\bigr)
-\sigma_0^p(g_1)\bigl(\sigma_u^{in}+\sigma_u^{q}+\sigma_u^{el}\bigr)
\over
\sigma_0^u\bigl((1+\delta_v)\sigma_0^u+\sigma_u^{in}+\sigma_u^{q}
+\sigma_u^{el}\bigr)
},
\label{rc02}
\end {equation}
 where $\de_v=\si^v_p/\si_0^p=\si^v_u/\si_0^u$. The Born and
inelastic radiative tail polarized parts of cross sections depend
on SF $g_1$, and in the last case the dependence is
non-trivial. So the equation \r{53} becomes functional one in
$g_1$. This functional equation transforms into a system considering the
extraction of $g_1$ in concrete binning over $x$
in $n$ kinematic points $x_i\; (i=1,...,n)$:
\begin {equation}
A^m_{1i} = {g_{1i}\over F_1} + \Delta A_1(g_{1j};j=1,...,n).
\label{54}
\end {equation}
Usually the iteration methods are used to solve such a system of
equations. The variant of iteration formula is ambiguous, but in
practice only two types are used. The first and most evident one
is to take for $k$-th step:
\begin {equation}
g^{(k)}_{1i} = F_1(A^m_{1i} - \Delta
A_1(g^{(k-1)}_{1j};j=1,...,n)).
\label{55}
\end {equation}

Another possibility to obtain the formulae for iteration procedure
arises when both Born and radiative correction cross section
are separated into spin-averaged and spin-dependent parts.
Then for the measured asymmetry we have
\begin {equation}
\displaystyle
A^m_1 =\frac 1D {\sigma_0^p+\sigma_1^p\over
\sigma_0^u+\sigma_1^u} =
{g_1/F_1 + \sigma_1^p/D\sigma_0^u\over
1 + \sigma_1^u/\sigma_0^u}.
\end{equation}
Thus we obtain the iteration formulae
\begin{equation}
g^{(k)}_{1i} = F_1\biggl[A^m_{1i}\biggl(1+\frac
{\sigma_1^u}{\sigma_0^u}\biggr)
 - {
\sigma_1^p(g^{(k-1)}_{1j};j=1,...,n)\over D \sigma_0^u} \biggr],
\label{56}
\end{equation}
where in right-hand side the dependence on $g_1$ is contained only
on the level of RC, but not on the Born level.

\begin{figure}
\unitlength 1mm
\begin{picture}(60,60)
\put(40,-20){
\epsfxsize=7cm
\epsfysize=7cm
\epsfbox{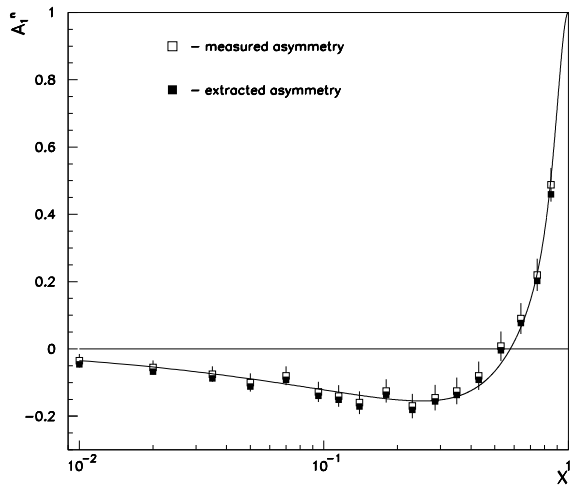}
}
\end{picture}
\caption{\protect\it
The results of iteration procedure for spin asymmetry
along with the fit constructed.
}
\label{Fgit}
\end{figure}

The 
above
presented 
formulae could be applied directly for
experimental data on hydrogen and deuteron. Three-parameter fit is
used for fitting of the spin asymmetry
\cite{SMC1,SMC2,SMC3,SMC4}:
\begin{equation}
\begin{array}{c}
A_1^p(x)=A+x^B(1-e^{-Cx}),
\\[3mm]
A_1^d(x)=(e^{-Ax}-1)(B^C-x^C).
\end{array}
\label{0014}
\end{equation}
For proton asymmetry the one- and two parameter fits
are also used:
\begin{equation}
A_1^p(x)=x^B, \qquad
A_1^p(x)=Ax^B.
\label{4439}
\end{equation}

For  the neutron asymmetry extracted from the deuteron
and helium-3 data the
nuclear corrections have to be taken into account:
\begin{equation}
A_1^n(x)={A^{D,^3\rm He}_{1}(x)-(1-f_d(x))P_pA_1^p(x)
\over f_d(x)P_n }.
\label{00d7}
\end{equation}

Dilution factors for both cases are given by the formulae
\begin{equation}
f_d(x)={1\over F_2^p/F_2^n+1}, \qquad
f_d(x)={1\over 2F_2^p/F_2^n+1}
\label{00060}
\end{equation}
respectively. The numbers $P_p$ and $P_n$
are effective nucleon polarizations in nucleus. For deuteron
target
$P_p=P_n=0.5-0.75\omega_d$, where $\omega_d$ is the D-state
probability ($\sim$5\%), and
$P_p$=-0.028 and $P_n$=0.86 for helium-3.

The application of procedure of RC for experiments on DIS with
polarized $^3$He and HERMES lepton beam with energies $E=27.5$ GeV
on fig.\ref{Fgit} are demonstrated.

Another important question carefully studied in all experiments on
lepton-nucleon scattering is contribution to systematical error
of measurement due
to radiative corrections. There are several sources for this uncertainty:
the RC procedure itself, dependence on SF models beyond measured region
and physical processes neglected in the standard RC procedure. Monte Carlo
approach for estimations of these effects is presented in \cite{AkuNag}
for the case of $^3$He polarized target.

\subsection{Semi-inclusive physics} \label{2.5}

We calculate the  radiative
corrections to data of semi-inclusive polarized experiments
\begin{equation}
 e(k_1) + N (p) \longrightarrow e'(k_2) + h(p_h) + X(p_x)
\end{equation}
when a hadron $h$ is detected in coincidence with the outgoing
lepton. In this case the cross section
depends on five kinematic variables which can be chosen as
\begin{equation}
 x,\;y;\,z,\;t,\;\phi_h,
\label{002}\end{equation}
where $x$ and $y$ are usual scaling variables, $z$ and $t$ are
defined via hadron momentum
\begin{equation}
t=(q-p_h)^2, \quad z=p_hp/pq, \quad q=k_1-k_2,
\label{005}\end{equation}
$\phi_h$ is an angle between planes (${\vec k_1},{\vec k_2}$) and
(${\vec q},{\vec p_h}$) in the rest frame ($p=(M,{\vec 0})$).
Instead of $t-$dependence we will also consider the cross section as a
function of transversal momentum $p_t$ defined in (A.8)
of \cite{AkuSor}.
 Also the following invariants will be used
\begin{eqnarray}
 M_x^2=p_x^2=(1-z)S_x+M^2+t,\;
V_{1,2}=2k_{1,2}p_h.
\end{eqnarray}

Notice that there are two tasks for this process. The first task
corresponds the description of semi-inclusive DIS in the frame of
quark-parton model. In this way the cross section
depends on three variables $x$, $y$ and $z$.
However for the real situation
the integration region is limited by the experimental cuts.
To deal with them on the angles of registered
hadron a  special procedure was developed and cross section
describing such situation is defined by five variables: three
standard $x$, $y$, $z$ and two additional  $p_t$, $\phi _h$.
The relation between these two cross sections can be established
as follows:
\begin{equation}
{d \sigma _{R}
\over dx dy dz}
 = \int
{1\over 2\pi}
 {d^3 k\over 2 k_0}
{d{\tilde z}\over dz}
dp_{t}
d\phi_h
\theta({\sin^2 \vartheta_{max}}- {\sin^2 \vartheta})
\theta({\sin^2 \vartheta}-
{\sin^2 \vartheta_{min}})
{d\sigma \over dx dy dz dp_{t} d\phi_H},
\label{sigRd}
\end{equation}
if $\theta $-function are removed.
Here  $\vartheta$ is the angle between the beam direction (${\vec
k_1}$) and hadron momentum in the lab frame.

Within quark-parton model
for the Born cross section of longitudinal polarized
semi-inclusive DIS we have
\begin{equation}
\sigma _{0}^s \equiv {d^{3}\sigma _{0} \over dxdydz} =
{2\pi \alpha ^{2} \over Sxy}[F^{u}_{0} \Sigma ^{+}(x,z) +
P_{L}P_{N}F^{p}_{0} \Sigma ^{-}(x,z)]
\label{71}
\end{equation}
where
\beq
F^{u}_{0} = 2( 1/y - 1 - \mu _{N}x ) + y, \qquad F^{p}_{0} = y -
2, \qquad \mu_N=M^2/S.
\eeq{si01}
The quantities
\begin{equation}
\Sigma ^{+(-)}(x,z) = \sum^{}_{q} e^{2}_{q} [f^{+}_{q}(x) \pm
f^{-}_{q}(x)] D^{H}_{q}(z)
\end{equation}
is the ordinary for QPM combination of the distribution  functions
$f^{+(-)}_{q}(x)$   for   the   quark   of  flower  $q$  polarized
(anti)parallel   to   the   nucleon   polarization,   and  of  the
fragmentation functions  $D^{h}_{q}(z)$ of  the quark  $q$ into
the
hadron $h$, $e_{q}$ being the quark charge in units of  elementary
charge.

As it was mentioned above there are three radiative tails give
contribution to DIS:inelastic, elastic and quasielastic (see
Fig \ref{gr}). However, in semi-inclusive process the
transfer energy
is above pion threshold, so in RC calculation
one have to take into account only $\sigma^{in}$ and
$\sigma^{v}$.
The lowest order QED correction was
calculated in
ref.\cite{Soroko1} and presented in \cite{P20} as the sum of
factorizing and non-factorizing infrared free parts
\begin{equation}
\sigma _{EM} \equiv {d^{3}\sigma _{EM}\over dxdydz} =
{\alpha \over \pi } \delta _{VR} \sigma _{0}^s+
\sigma ^{F}_{R},
\end{equation}
where the first one
is the sum of infrared divergence separated from radiative tail
and the additional virtual photon contribution:
$\delta _{VR} = \delta _{V} + \delta ^{IR}_{R}$.

The finite part of \r{71} has the form
\begin{equation}
\begin{array}{l}
\displaystyle
\sigma ^{F}_{R}
= {2\alpha ^{3}\over S}
\int\limits_{t_{1d}}^{t_{1u}}dt_1
\int\limits_{t_{2d}}^{t_{2u}}dt_2
 \{
{y^{2}\over x^{2}_{t}y_{t}} [F^{u}_{R}\Sigma^{+}
(\tilde{x},\tilde{z}) + P_{L}P_{N}F^{p}_{R}\Sigma ^{-}
(\tilde{x},\tilde{z})] - \\ [0.5cm]
\displaystyle \qquad  \qquad
- {\theta (t_{1i}-t_{1})\over xy} [F^{u}_{IR}\Sigma ^{+}(x,z) +
P_{L}P_{N}F^{p}_{IR}\Sigma ^{-}(x,z)] \},
\end{array}
\label{sigFR}
\end{equation}

The definition of the rest quantities can be found in \cite{P20}

In order to write five-dimensional cross section
it is useful to take the some scalar  products with
$p_h$ in the form
\begin{eqnarray}\label{009}
 &&
\frac{1}{2}
V_{1,2}= k_{1,2}p_h =  a^{1,2}+b\cos\phi_h\;\;
\\
&&
\frac{1}{2}\mu R=
kp_h  =
R(a^{k}+b^k(\cos\phi_h\cos\phi_k+\sin\phi_h\sin\phi_k)).
\nonumber\end{eqnarray}
where $\mu = kp_h/kp$ and
$\phi_k$ is the
rest frame
 angle between planes (${\vec k_1},{\vec k_2}$) and
(${\vec q},{\vec k}$).
We will use $a^{\pm}=a^2\pm a^1$.
The explicit expressions for coefficients
are given in \cite{AkuSor}.

As a result the five-dimensional cross section of unpolarized DIS
on the Born level has the following dependence on $\phi_h$:
\begin{equation}
\sigma_0 = {d\sigma_0\over dxdydzdp_t^2d\phi_h}=
\frac{N}{Q^4}
(A+\cos\phi_h A^c+\cos
2\phi_h A^{cc}),
\label{010}\end{equation}
where $N=\alpha^2 y S_x/\sqrt{\lambda_Q}$.
The coefficients $A$, $A^c$ and $A^{cc}$ do not depend on
$\phi_h$ more and they have the form
\begin{eqnarray}
A&=&
 2Q^2{\cal H}_1
+ (SX - M^2Q^2){\cal H}_{2}
+ ( 4a^1a^2 + 2b^2-M_h^2Q^2){\cal H}_{3}
\nonumber\\&&
+ (2Xa^1+2Sa^2 - zS_xQ^2){\cal H}_{4},
\nonumber\\
A^c&=& 2b(2a^+{\cal H}_{3} + S_p{\cal H}_{4}),
\nonumber\\
A^{cc}&=& 2b^2{\cal H}_{3}.
\end{eqnarray}
The model for structure functions ${\cal H}_{i}$ can be constructed
on basis of results
of the paper Mulders and Tangerman \cite{Muld1}. Keeping only
the leading twist contribution  we have for structure functions
\begin{equation}
\begin{array}{ll}
\label{model}
\displaystyle
{\cal H}_{1} =\sum_q e_q^2 f_q(x)D_q {\cal G}
,
&\displaystyle
{\cal H}_{2} =-\frac{p_t^2+m_h^2}{M^2E_h^2}\sum_q e_q^2 f_q(x)D_q {\cal  G}
,\nonumber\\[0.3cm]
\displaystyle
{\cal H}_{3} =0
&\displaystyle
{\cal H}_{4} =\frac{1}{ME_h}\sum_q e_q^2 f_q(x)D_q {\cal G},
\end{array}
\end{equation}
and
\begin{equation}
{\cal G}={\cal G}_1=b\exp(-bp_t^2)
,
\label{calG}\end{equation}
where $b=R^2/z^2$ is a slope parameter and $R$ is a parameter of the model.
Alternatively power fit is also considered in \cite{AkuSor}.

When integrated over the kinematic variables $\phi_h$ and $p_t$ the
cross section (\ref{010}) coincides with the unpolarized part
of semi-inclusive cross section calculated within quark parton model (\ref{71}).

The cross section that  takes into account radiative effects can
be written as
\begin {equation}
\sigma _{obs} = \sigma _0 e^{\delta_{inf}}
(1+ \delta_{VR}+\delta_{vac})+\sigma_{F}.
\label{eq11}
\end {equation}
Here the corrections
 $\delta_{inf}$  and   $\delta_{vac}$  come from  radiation of soft
photons \cite{Sh} and effects of vacuum polarization\footnote{There are explicit
formulae for leptonic contribution to vacuum polarization effect
(see \protect\cite{ASh} for example) and parameterization of hadronic one
\protect\cite{delvac}.}.
The correction  $\delta_{VR}$ is infrared free sum of factorized
parts of real and virtual photon radiation. These quantities are
given by the following expressions
\begin{eqnarray}        \label{deltas}
\delta_{VR} &=&\frac{\alpha}{\pi}
\biggl(\frac{3}{2}l_m\!-\!2\!-\frac{1}{2}\ln^2\frac{X'}{S'}+{\rm
Li}_2
\frac{S'X'-Q^2p_x^2}{S'X'}-\frac{\pi^2}6\biggr)\!,
\nonumber\\
\delta _{inf}&=& \frac{\alpha}{\pi}
(l_m-1)\ln\frac{(p_x^2-(M+m_{\pi})^2)^2}{S'X'},
\\
\delta _{vac}&=& \delta_{vac}^{lept}+\delta_{vac}^{hadr},\nonumber
\end{eqnarray}
where $S'=X+Q^2-V_2$, $X'=S-Q^2-V_1$, $l_m=\ln Q^2/m^2$ and ${\rm
Li}_2$
is Spence function or dilogarithm.

The contribution of radiative tail has the standard form
(see eq.(\ref{eq25}))
\begin{eqnarray}\label{eq252}
\sigma _F &=& -{\alpha N \over 2 \pi}
 \int\limits^{2\pi }_{0}d\phi_k
 \int\limits^{\tau_{max}}_{\tau_{min}}d\tau
 \sum^{4}_{i=1}
 \sum^{3}_{j=1} \theta
_{ij}(\tau,\phi_k) 
\int\limits^{R_{max}}_{0}
dR R^{j-2} \biggl[
{{\cal H}_i\over (Q^2+R\tau)^{2}}-
\delta_j{{\cal H}_i^0\over Q^4}\biggr] .
\end{eqnarray}
Here $2M^2\tau_{max,min}=S_x\pm \sqrt{\lambda_Q}$ and $R_{max}
=(M_x^2-(M+m_{\pi})^2)/(1+\tau-\mu)$, $\delta_j$=1 for
$j$=1 and $\delta_j$=0 otherwise. The explicit formulae for
functions
 $\theta(\tau,\phi_k)$ can be found in \cite{AkuSor}.
The structure functions ${\cal H}_{i}^0 $ have to be calculated for Born
kinematics, but ${\cal H}_{i}$ is calculated in terms of tilde
variables introducing by
\begin{eqnarray}\label{til}
 {\tilde Q^2}  &=&  (q-k)^2  =  Q^2+R\tau, \nonumber\\
 {\tilde W^2}  &=&  (p+q-k)^2  =  W^2-R(1+\tau), \nonumber\\
 {\tilde t}  &=&  (q-k-p_h)^2  = t+R(\tau-\mu), \nonumber\\
 {\tilde M_x^2}  &=&  {\tilde p_x}^2  = M_x^2+R(1+\tau-\mu).
\end{eqnarray}
\markright{ }
\subsection{Higher order effects}
\markright{ }

The dominant contribution to RC can be obtained by the
method of leading logarithms. Being based on the fermion mass
factorization this approximation allows us to calculate
corrections $\sim (\alpha \log (Q^2/m^2_f))^n$. The
below formulae were obtained in \cite{LO} by the method
which is presented in
\cite{KurSpi}.

The contribution of inelastic tail  $\sigma ^{in}_2$ together with
loop effects
$\sigma ^{v}_2$
\begin{equation}
\sigma ^{in}_2+\sigma^v_2= \delta^{in}_{2}\sigma_0
+\sigma _{Vk_1}^{in}
+\sigma _{Vk_2}^{in}
+\sigma _{k_1k_1}^{in}
+\sigma _{k_2k_2}^{in}+\sigma
_{k_1k_2}^{in}
+\sigma _{lk_1}^{in}+\sigma _{lk_2}^{in}
+\sigma _{fk_1}^{in}+\sigma _{fk_2}^{in}
.
\label{0054}
\end{equation}

The factorized part is
\begin{equation}
\delta^{in}_2=\frac
{\alpha^2}{4\pi^2}(3\delta^2(Q^2)+
2l_m\delta_{sp}\delta(Q^2)
+\frac 1 2 l_m^2
\bigl[ \delta_{sp}^2
+4{\rm Li}_2(1-z_2)
+12{\rm Li}_2(1-z_1)-\frac 8 3
{\pi}^2\bigr])
,
\label{00551}
\end{equation}
$l_m=\log Q^2/m^2$, $\delta _{sp}=2\log((1-z_1)(1-z_2))+3$,
and $\delta(Q^2)=\delta^l_{vac}+\delta^h_{vac}$.

The contribution from the vacuum polarization if it coincides with
real photon radiation has the form
\begin{equation}
\sigma _{Vk_1,Vk_2}^{in}=\frac {\alpha^2}{2\pi^2}l_m
\int\limits_{z_{1,2}}^{1}
\frac{dz}{1-z}
\left[(1+z^2)\delta(t_{x,s})\sigma _{s,p}-2\delta(Q^2)\sigma_0
\right],
\label{l004}
\end{equation}
where $t_x=zQ^2$ and $t_s=Q^2/z$.

Next three terms correspond to the cases when two radiated photons
are collinear to an incident electron ($\sigma_{k_1k_1}$),
an outgoing electron ($\sigma_{k_2k_2}$)
\begin{equation}
\begin{array}{l}
\displaystyle
\sigma _{k_1k_1,k_2k_2}^{in}=\frac {\alpha^2}{8\pi ^2}
l^2_m \int \limits _{z_{1,2}}^1 dz\Biggl [ \frac 2{1-z}
(2\log [(1-z)(1-z_{1,2}(z))]-\ln z+3)\times
\\[5mm] \qquad \qquad
\displaystyle
\times ((1+z^2)\sigma
_{k_{1,2}}-2\sigma _0)
+((1+z)\ln z-2(1-z))\sigma _{k_{1,2}}\Biggl],
\end{array}
\label{0056}
\end{equation}
or when one photon is radiated in incident electron direction and
the other in the outgoing electron direction ($\sigma_{k_1k_2}$):
\begin{equation}
\begin{array}{l}
\displaystyle
\sigma _{k_1k_2}^{in}=\frac {\alpha^2 }{4\pi^2}l_m^2
 \int \limits _{z_1}^1 \frac {dz'}{1-z'}
\int\limits _{z_2(z')}^1\frac {dz}{1-z}
\Biggl [(1+z^2)(1+z'^2)\sigma _{k_1k_2}
-2(1+z'^2)\sigma _{k_1}
\\[5mm]
\displaystyle
\qquad \qquad
-2(1+z^2)\sigma _{k_2}
+4\sigma _{0}
\Biggl] .
\end{array}
\label{00572}
\end{equation}
Here
\begin{equation}
\begin{array}{c}
\displaystyle
\sigma _{k_1k_2}=y\sigma_0(z'S,X/z,z'Q^2/z)/(z(zz'-1+y)),
\\[5mm]
\displaystyle
z_1(z)=(1-y)/(z-xy),
z_2(z)=(1-y+zxy)/z.
\end{array}
\label{997}
\end{equation}

There are two channels (singlet and non-singlet) of the
fermion pair production that give a contribution to $\alpha^2$ of
order RC.
The singlet channel
\begin{equation}
\begin{array}{l}
\displaystyle
\sigma _{lk_1,lk_2}^{in}= \frac {\alpha^2 }{8\pi^2 }
l^2_m \int \limits _{z_{1,2}}^{1-4m_lM/S} dz
(2(1+z)\log z+1-z+\frac 4 3(1-z^3)z)\sigma _{k_{1,2}}
\end{array}
\label{0058}
\end{equation}
corresponds to the case when the incident and the outgoing lepton
as well as the leptons of the unregistered pair belong to
different leptonic lines connected by an additional virtual photon.

The rest non-singlet part
\begin{equation}
\sigma _{fk_1,fk_2}^{in}=\frac {\alpha^2 }{12\pi^2}\sum _f
\log ^2 \frac{Q^2}{m_f^2} \int \limits
_{z_{1,2}}^{1-4m_fM/S} dz \frac{(1+z^2)}{(1-z)}\sigma _{k_{1,2}}
\label{0060}
\end{equation}
arises from the two-lepton decay of an additional virtual photon.

The main contribution to the second order elastic and quasielastic
radiative tail arises when the additional radiated photon is
collinear to the incident or outgoing fermion line:
\barr{c}
\sigma^{el}_2=\od {\alpha }{2\pi}l_m\delta_{sp}\sigma^{el}_1
+\sigma^{el}_{k_1t}+\sigma^{el}_{k_2t},
\earr{0118}
where
\barr{c}
\sigma^{el}_{k_{1,2}t}=\od {\alpha }{2\pi}l_m\int
\limits_{z_{1,2}}^1
dz\od {(1+z^2)\sigma^{el}_{k_{1,2}}-2\sigma^{el}_{1}}{1-z}.
\earr{0119}
The quantities $\sigma^{el}_{k_{1,2}}$ are obtained in the terms
of approximate elastic radiative tail
$\sigma^{el}_{1}=\sigma^{el}_{1}(x,y,S)$:
\barr{l}
\sigma _{k_1}^{el}=y\sigma _1^{el}( xyz/(z+y-1), (z+y-1)/z,
zS)/(z-1+y),
\di5
\sigma _{k_2}^{el}=y\sigma _1^{el}( xy/(z+y-1), (z+y-1)/z,
S)/z(z-1+y).
\earr{ddfe}

The numerical analysis shown that correction $\sim \alpha^2$ to the
considered asymmetry is about some per cent from the full RC and
has a maximum near kinematic borders of the considered
experiments.

However it is necessary to remind that correction  $\sim \alpha^2 $
has been calculated within LO approximation while even
ultrarelativistic approximation for radiative tails from elastic
and quasielastic peaks leads to high corrections. Besides, the
approximation $\sim \alpha^2 \log^2(Q^2/m^2)$ gives rather good
results for the factorizad part of correction but it produce
poorer ones for the remaining components which make the dominant
contribution to the asymmetry in polarized particle scattering.
Hence, a more accurate calculation is required for
more detail analysis of $\sim \alpha^2 $ correction.

\subsection{FORTRAN codes POLRAD 2.0 and HAPRAD}
\begin{figure}
\unitlength 1mm
\begin{tabular}{cc}
\begin{picture}(60,60)
\put(0,-22){
\epsfxsize=7cm
\epsfysize=7cm
\epsfbox{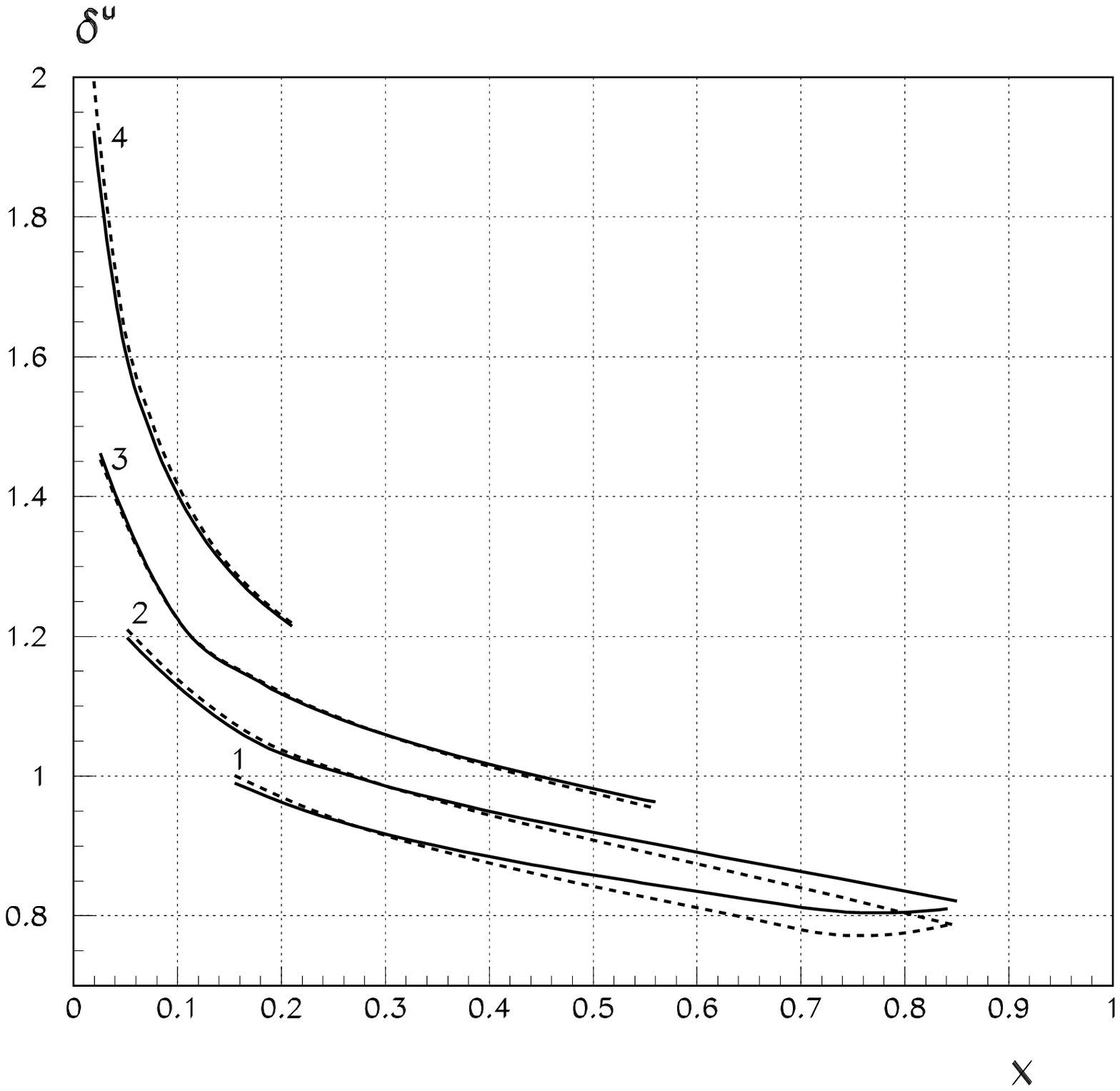}
}
\put(37,-7){{\rm a)}}
\end{picture}
&
\begin{picture}(60,60)
\put(20,-22){
\epsfxsize=7cm
\epsfysize=7cm
\epsfbox{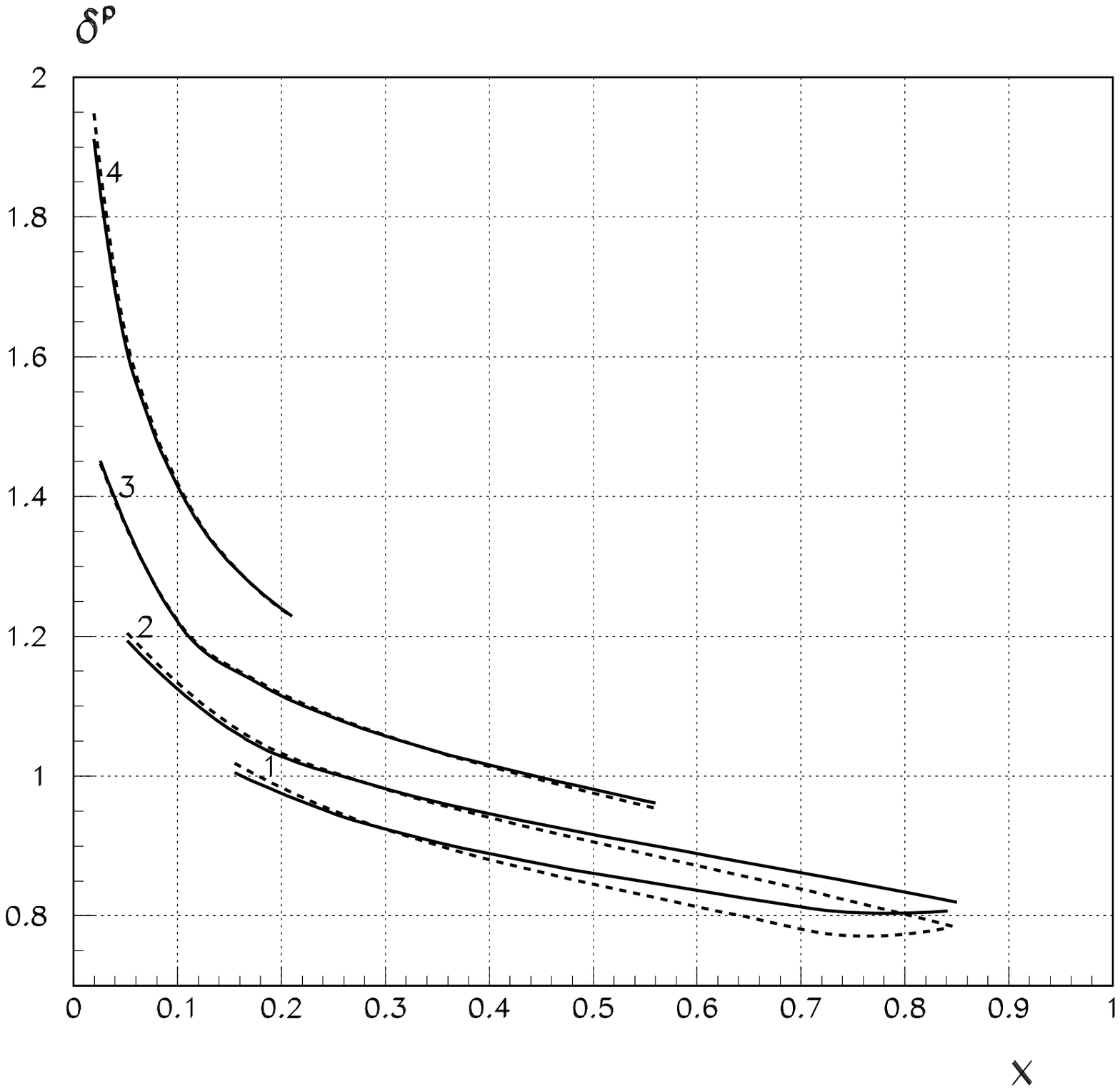}
}
\put(55,-7){{\rm b)}}
\end{picture}
\end{tabular}
\vspace{5mm}
\caption{\protect\it
$x-$dependence of the ratio of the total cross section to the
Born
on with (dashed curves) and without (solid curves) the second
order effects for electron-proton DIS with HERMES kinematics at
the initial lepton energy of
$E_e=27.5 GeV$ for
1) $y=0.1$, 2) $y=0.3$, 3) $y=0.6$, 4) $y=0.8$;
a)the correction to unpolarized part of cross section;
b)the correction to polarized part of cross section.
}
\label{ap1}
\end{figure}
Using the results presented in the previous subsections two
FORTRAN code
POLRAD 2.0 \cite{P20} and HAPRAD \cite{AkuSor} have been created.
Here some results obtained by these codes are presented.

The program POLRAD 2.0 consists of two patches. The first one
which itself is called POLRAD describes RC to inclusive
DIS. In order to estimate some radiative effects by this patch
let us separate cross section into spin-averaged and
spin-dependent parts
\begin {equation}
\sigma _{0,1,2}=\sigma ^u_{0,1,2}+P_N\sigma ^p_{0,1,2}.
\label{rrt}
\end {equation}
and consider the following quantities:
\begin {equation}
\delta^{u,p} _{1}=\od{\sigma ^{u,p}_{0}+\sigma ^{u,p}_{1}}
{\sigma ^{u,p}_{0}},\;\;\;\;\;\;\;\;\;
\delta^{u,p} _{2}=\od{\sigma ^{u,p}_{0}+\sigma ^{u,p}_{1}
+\sigma ^{u,p}_{2}}{\sigma ^{u,p}_{0}},
\label{dpu}
\end {equation}
which are characterized the ratio of the total unpolarized and
polarized part of the cross section to the Born one (with and
without second order effect). The dependence of these quantities
on scaling variable in HERMES kinematics is presented in fig.
\ref{ap1}.

\begin{figure}
\vspace{30mm}
\unitlength 1mm
\begin{picture}(60,60)
\put(0,-22){
\epsfxsize=13.5cm
\epsfysize=13.5cm
\epsfbox{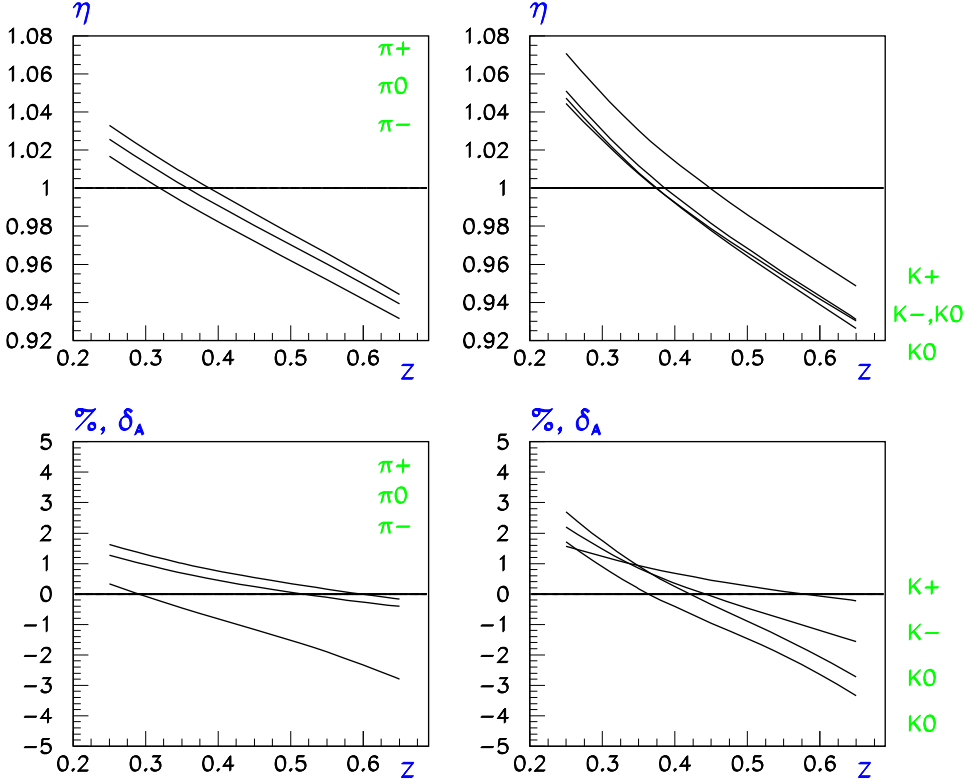}
}
\end{picture}
\vspace{5mm}
\caption{
\protect\it
The radiative correction factor to semi-inclusive
cross sections ($\eta=\sigma / \sigma_0$, upper plots) and to
polarization asymmetry
($\delta _A=(A^{obs}-A^{born})/A^{born}$, lower plots) versus z. The cross
sections were
integrated over other kinematic variables within HERMES cuts
\cite{Manuella}.
}
\label{ex3}
\end{figure}

The second patch of POLRAD 2.0 which is called as SIRAD allows us
to calculate RC to three-dimension cross
section $d\sigma /dxdydz $ of semi-inclusive DIS. Fig. \ref{ex3} shows
the ratio  $\eta=\sigma^{obs}/\sigma^{born}$ and the
relative
correction to asymmetry
$\delta_A=(A^{obs}-A^{born})/A^{born}$ for $K$ and $\pi
$ mesonproduction.

To estimate the RC
to five-dimensional cross section
$d^5\sigma/dxdydzdp_t^2d\theta_h$
the special FORTRAN code HAPRAD was developed.
Now we show that numerical results for RC
to $d\sigma/dxdydz$
reproduced by these two codes which coincide with a good accuracy.
It
can be seen in the Table \ref{tab1} where we represent
RC to the cross
section as it follows from the runs of the codes POLRAD 2.0 and HAPRAD.
Since HAPRAD allows us to take into account the kinematic cuts
and to
use different models for $p_t^2$-slope,
three fits for $p_t^2$-distribution
and
the cases with and without experimental cuts were considered.  The
first fit for $p_t^2$-slope is defined in (\ref{calG}),  while
the second and third ones are our
fits of experimental data \cite{EMCS} using exponential
(${\cal G}'_1={\cal G}_1$ at $b=a/z$) and power functional
forms
\begin{equation}
{\cal G}={\cal G}_2=\biggl[{1\over a+bz+p_t^2}\biggr]^{c+dz}.
\label{calG2}\end{equation}
As the kinematic
cuts on
$\phi_h$ and $p_t$ of the measured hadron we took HERMES geometrical
cuts \cite{HERMESsi}.
 We can conclude from this analysis that
 neither important differences between SIRAD and HAPRAD results,
if exponential model for $p_t^2$-distribution is used,   nor
dependence on
slope parameter model and applying of geometrical cuts are found.
However RC takes a negative shift in the model
based on power functional form (\ref{calG2}). As it is shown in
\cite{AkuSor}
 RC depends on steepness of $p_t^2$ distribution. It is a
reason why models like $\delta(p_t^2)$ (QPM, POLRAD 2.0) and
(\ref{calG}) give larger RC. Within practical RC procedure
in concrete measurement of $d\sigma/dxdydz$ the model can be
fixed only if the information about $p_t^2$-distribution is
additionally considered.

\begin{table*}[!t]
\begin{center}
\begin{tabular}{c|c|c|c|c|c|c|c|c|c|c}\hline
&&&&SIRAD&\multicolumn{6}{c}{HAPRAD}\\ \cline{5-11}
$x$ & $y$ & $Q^2$  & $z$ &  &  \multicolumn{3}{c|}{without  cuts}&
  \multicolumn{3}{c}{with cuts}
 \\ \cline{6-11}
 &&GeV$^2$&&&${\cal G}_1$&${\cal G}'_1$&${\cal G}_2$&${\cal G}_1$
 &${\cal G}'_1$&${\cal G}_2$
\\
\hline
  \hline
0.038 & 0.677 & 1.33 & 0.25 & 1.029 &1.033&1.024&0.982& 1.041&1.025&0.985\\ 
0.062 & 0.567 & 1.82 & 0.35 & 0.996 &0.989&0.989&0.947& 0.989&0.980&0.951\\ 
0.092 & 0.529 & 2.52 & 0.45 & 0.970 &0.961&0.961&0.934& 0.961&0.956&0.936\\ 
0.131 & 0.499 & 3.38 & 0.55 & 0.945 &0.936&0.933&0.912& 0.934&0.931&0.906\\ 
0.198 & 0.476 & 4.88 & 0.65 & 0.918 &0.902&0.902&0.889& 0.897&0.897&0.881\\ \hline
\end{tabular}
\end{center}
\caption{\it
The results for RC factors to three dimensional semi-inclusive
cross section obtained by FORTRAN codes SIRAD and HAPRAD (see
text for further explanations). Kinematic points are taken from the
Table 1 in ref.\cite{Manuella}.
}
\label{tab1}
\end{table*}

\subsection{Monte Carlo generator RADGEN 1.0}

In this subsection we present a Monte Carlo generator RADGEN 1.0
\cite{RAD} for the events with a possible radiation of a real
photon in DIS on polarized and unpolarized targets.
The
events are generated in accordance with their contribution to the
observed total cross section given by
\begin{equation}
\sigma_{obs}=\sigma_{non-rad}(\Delta)+\sigma_{in}(\Delta)
+\sigma_{q}+\sigma_{el}.
\label{addrc00}\end{equation}
The first term $\sigma_{non-rad}(\Delta)$
contains not only the contribution from the
Born process but the contributions from loop corrections
($\sigma_v$) and from multiple soft photon production with a total energy
not exceeding a cut-off parameter $\Delta$. In handling the radiative
corrections two approaches have been used, the one developed by Mo and
Tsai \cite{MoTsai,Tsai} and the other given by Bardin and Shumeiko
\cite{BSh}.

Kinematics of the Born (or one photon exchange) process
is completely defined by the scattering angle $\theta$
and the energy $E'$ of the scattered lepton, two variables
which are usually measured.
All other inclusive
kinematic variables can be expressed in the terms of $\theta$ and
$E'$. Having in mind the laboratory frame, i.e. the frame with the target
nucleus at rest the relevant variables are
\begin{equation}\begin{array}{c}\displaystyle
Q^2=4EE'\sin^2\frac{\theta}{2}, \qquad
\nu=E-E', \qquad
x={Q^2\over 2M\nu},
\\[0.5cm]\displaystyle
y=\frac{\nu}{E},\qquad
W^2=Q^2(1/x-1)+M^2,
\end{array}\label{0010}\end{equation}
where $E$ is the beam energy.

The event registered in the detector with certain values of
$E'$ and $\theta$ for the scattered lepton can either be a
non-radiative  or a radiative event, i.e. an event containing a real hard
radiated photon.
For the radiative event there are
three additional variables necessary to fix the kinematics of
the radiated real photon apart from $E'$ and $\theta$.
A possible choice is the
photon energy $E_{\gamma}$ and the two angles $\theta_{\gamma}$ and
$\phi_{\gamma}$, where $\theta_{\gamma}$
is the angle between the real and the virtual photon momenta
$\vec k$ and $\vec q=\vec k_1-\vec k_2$ and
$\phi_{\gamma}$ is the angle between the planes defined by the
momenta ($\vec k_1$, $\vec k_2$) and ($\vec k$, $\vec q$).
For the events with the radiation of a real photon the kinematic
variables describing the virtual photon and which is used to
generate
the hadronic final state is derived from eq.(\ref{0010}) because
the substitution  $q \rightarrow q-k$ has to be made in their definition.
The variables obtained after this
 substitution will be referred to as 'true' ones:
\begin{equation}
\begin{array}{ll}\displaystyle
W^2_{true}=W^2-2E_{\gamma}(\nu +
M-\sqrt{\nu^2+Q^2}\cos\theta_{\gamma}),
&\displaystyle
\nu_{true}=\nu-E_{\gamma},
\\[0.5cm] \displaystyle
Q^2_{true}=Q^2+2E_{\gamma}(\nu
-\sqrt{\nu^2+Q^2}\cos\theta_{\gamma}),
&\displaystyle
x_{true}={Q^2_{true} \over 2M\nu_{true}}.
\end{array}
\label{true}
\end{equation}
For non-radiative events the true kinematics exactly coincide with
(\ref{0010}).

In case of the elastic and quasielastic radiative processes the mass
squared of the hadronic final state is fixed imposing additional
constraints on the true kinematic variables. Indeed the following
ranges are allowed:
\begin{equation}
\left\{
\begin{array}{cl}
\displaystyle x\leq x_{true}\leq 1 & \displaystyle{\rm for}\;\;
\sigma_{in}\\[0.5cm]
\displaystyle x_{true}= 1 & \displaystyle{\rm for}\;\;
\sigma_{q}\\[0.5cm]
\displaystyle x_{true}=M_A/M & \displaystyle{\rm for}\;\;
\sigma_{el}
\end{array}
\qquad
Q^2_{min}\leq Q^2_{true}\leq Q^2_{max}
\right.
\label{0011}\end{equation}
where
\begin{equation}
Q^2_{max,min}=Q^2{2(1-x_r)(1\pm \sqrt{1+\gamma^2})
+\gamma^2 \over \gamma^2+4x_r(1-x_r)}
\label{0012}\end{equation}
and
\begin{equation}
\gamma^2=4M^2x^2/Q^2, \qquad x_r=x/x_{true}.
\label{0013}\end{equation}

The Monte-Carlo procedure for the generation of events with a possible
photon radiation
is the following:

\noindent
The procedure starts off with the generation of the kinematics of
the scattered
lepton and the calculation of an event weight from these kinematics.
Then the appropriate scattering channel (non-radiative; elastic,
quasielastic or inelastic radiative tail) has to be chosen according
to their contribution to the total observed cross section (see
(\ref{addrc00})).
If a radiative channel is selected the radiated photon has to be
generated and the values of the kinematic variables have to be
re-computed to obtain the true values. For each event the
weight has to be recalculated. The new weight is defined as the
ratio of the radiatively corrected and the Born cross section. After
this recalculation the weighted sum of all events (generated originally
in accordance with the Born cross section) gives the observed cross
section.

The recalculation of the weight requires the knowledge of the cross
section integrated over the photon momentum. This is done differently
in two theoretical approaches. In the Mo-Tsai approach an additional
parameter
$\Delta$ dividing the integration region into a soft and a
hard photonic part is introduced. There is no such parameter
in the Bardin-Shumeiko approach.
However, in both approaches a minimal photon energy $ E_{min}^{\gamma}$
is adapted in generating a radiated photon. If the photon energy
is
above this value the kinematics of the photon is calculated and the event
becomes a radiative one. The actual value of $ E_{min}^{\gamma}$ depends
on the aim of the photon generator. In general photons should be detected
in the calorimeter. So the energy threshold of the calorimeter sets the
value for $ E_{min}^{\gamma}$.

The unpolarized generator is based on the FORTRAN code FERRAD35
\cite{FERRAD}. The code
calculates the radiative correction to deep inelastic scattering
of unpolarized particles in accordance with the analytical formulae
given by Mo and Tsai \cite{MoTsai,Tsai}.

The result for the lowest order radiative correction of the cross
section is
\begin{equation}
\sigma  = \delta_{R}(\Delta)(1+\delta_{vert}+\delta_{vac}+\delta_{sm})\sigma_{1\gamma}
+ \sigma_{el}+ \sigma_{q}+ \sigma_{in}(\Delta)
,
\label{001}\end{equation}
where $\sigma_{1\gamma}$ is the one photon exchange Born cross section and
$\delta_{vac}$, $\delta_{vert}$, and $\delta_{sm}$ are
corrections due to vacuum polarization by electron and muon pairs,
vertex corrections and
residuum of the cancellation of infrared divergent terms independent of
$\Delta$ (see \cite{RAD} for detail).

The cross sections $\sigma_{el}$, $\sigma_{q}$, and $\sigma_{in}$
are the contributions from radiative processes (i.e. including real photon
radiation) for
elastic, quasielastic and deep inelastic scattering
(see \cite{MoTsai, Tsai,RAD} for detail).

An artificial parameter $\Delta$ had to be introduced to divide the
integration region over the photon energy into two parts, the soft and the
hard energy regions. The hard energy region $\sigma_{in}(\Delta)$ can be
calculated without any approximations. The soft photon part is calculated
for photon energies approaching zero. After cancellation of the infrared
divergences and a resummation of soft multiphoton effects the correction
factor is given by
\begin{equation}
\delta_R(\Delta)=\exp\left[-
\frac{\alpha}{\pi}\left(\ln \frac{E}{\Delta}+\ln \frac{E'}{\Delta}\right)
\left(\ln \frac{Q^2}{m^2}-1\right)\right].
\label{0022}\end{equation}

The polarized generator is constructed utilizing the FORTRAN code
POLRAD 2.0 \cite{P20,ASh} that was presented above.
The cross
section formulae (\ref{eq1}-\ref{eq23q}) can be rewritten into a form
similar to eq.(\ref{001}):

\begin{equation}
\sigma  = \delta_{R}(
E^{\gamma}_{min})(1+\delta_{vert}+\delta_{vac}+\delta_{sm})\sigma_{1\gamma}
+\sigma_{add}( E^{\gamma}_{min})
+ \sigma_{el}+ \sigma_{q}+ \sigma_{in}( E^{\gamma}_{min}).
\label{007}\end{equation}

In the ultrarelativistic approximation ($Q^2\gg m^2$) the
corrections $\delta_{vert}$, $\delta_{vac}$, and $\delta_{sm}$ take the
same form as in eq.(\ref{001}) \footnote{It should be noted that
contributions from $\tau$-leptons and quark loops were included into
$\protect\delta_{vac}$ \protect\cite{ASh}.}.
Since this code is based on the method of covariant cancellation
of infrared divergences developed by Bardin and Shumeiko
\cite{BSh} the final formula for the cross section free of
infrared divergences does not include any artificial parameter like
$\Delta$. However, $ E^{\gamma}_{min}$ has to be introduced if the code
is used considering the generation of real radiated photons.
As a result the formula (\ref{eq25}) has to be
rewritten in the following way
\begin{equation}
\label{b001}
\sigma _{in}= - {\alpha ^{3}y}\int\limits^{\tau_{max}}_{\tau_{min}}
d\tau\sum^{4}_{i=1}
 \sum^{k_{i}}_{j=1} \theta _{ij}(\tau)\int\limits^{R_{max}}
 _{R_{min}} dR
 {R^{j-2}\over (Q^2+R\tau)^{2}}\Im_{i}(R,\tau),
\end{equation}
while the elastic and quasielastic radiative tails were defined
above by eq.(\ref{eq23}) and (\ref{eq23q}) respectively. The
lower integration limits is determined by
$R_{min}=2M E^{\gamma }_{min}$.

Apart from the usual ultrarelativistic approximation another one was made
when
the eqs. (\ref{001}) and (\ref{002}) were obtained, namely the photon
energy was considered to be small $ E^{\gamma} \ll E,E'$ in the region
$E^{\gamma}<
E^{\gamma}_{min}$.
In POLRAD, the term $\sigma_{add}(E_{min}^{\gamma})$
is added to take the difference of this approximation to the exact
formula
into account:
\begin{eqnarray}\label{add}
\sigma_{add}( E^{\gamma}_{min}) &=& -{\alpha ^{3}y}
 \int\limits^{\tau_{max}}_{\tau_{min}}d\tau \sum^{4}_{i=1}
 \left\{ \theta _{i1}(\tau)
\int\limits^{2M E^{\gamma}_{min}}_{0}{dR\over R} \right.
\left[ {\Im _{i}(R,\tau )\over (Q^2+R\tau )^2}-{\Im
_{i}(0,0)\over Q^4}\right]
\nonumber\\
&&\qquad \qquad\qquad\qquad\qquad\left.
+ \sum^{k_{i}}_{j=2} \theta _{ij}(\tau)\int\limits^{2M
E^{\gamma}_{min}}_{0}
dR {R^{j-2}\over (Q^2+R\tau )^{2}}\Im _{i}(R,\tau)  \right\}.
\end{eqnarray}
\begin{figure}[t]
\unitlength 1mm
\begin{picture}(160,130)
\put(10,-10){
\epsfxsize=15cm
\epsfysize=15cm
\epsfbox{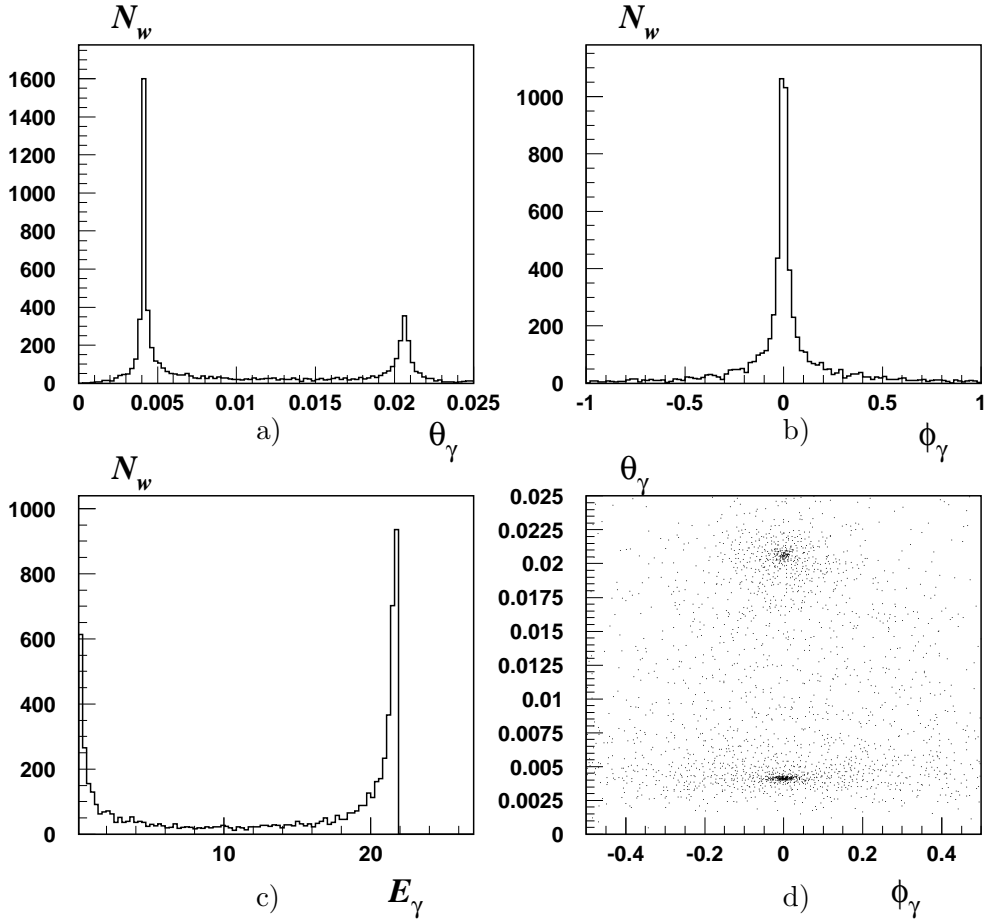}
}
\put(50,65){{\rm \small a)}}
\put(120,65){{\rm \small b)}}
\put(50,3){{\rm \small c)}}
\put(120,3){{\rm \small d)}}
\end{picture}
\caption{\protect\it \label{disg}
The distribution of the radiation angles $\theta_{\gamma}$ a) and
$\phi_{\gamma}$ b) and of the energy c) of the
radiated photon for $x=0.1$ and $y=0.8$. The two-dimensional
distribution d) shows $\theta_{\gamma}$ vs $\phi_{\gamma}$.}
\end{figure}
Therefore, the results obtained with the help of POLRAD are
independent of the threshold parameter $E_{min}^{\gamma}$ by
construction.

To estimate the value of radiative effects numerically
Monte-Carlo event
samples of 100k events each have been generated
for a $^3$He target and kinematic cuts relevant for the
HERMES experiment \cite{HERMES1} have been applied
($Q^2>$1 GeV$^2$, $W^2>$4 GeV$^2$, $y<$0.85, 0.037$<\theta<$0.14$rad$).
The distributions of the radiation angles
$\theta_{\gamma}$ and $\phi_{\gamma}$ as defined in
the so-called Tsai-system in which the $z$-axis is along the
direction of the virtual photon and the $y$-axis is normal to the
scattering plane (see ref.\cite{MoTsai, Tsai}) and of the energy
$E_{\gamma}$ are displayed in fig.\ref{disg}. These distributions are
generated for a certain point in the kinematic plane of the scattered
lepton, i.e. for $x$ = 0.1 and $y$ = 0.8.
The two peaks seen in the $\theta_{\gamma}$ distribution of fig.\ref{disg}
correspond to the $s$- and $p$-peak emerging from collinear radiation
along the direction of the incoming and outgoing lepton, respectively.

\subsection{Electroweak correction}

It is natural, that data processing of modern experiments on DIS requires
correct account of RC.  The above presented expressions for RC to
polarized DIS within QED is successfully used for fixed target
experiments.

However when the polarization DIS
experiments at collider will be possible \cite{herap},
we cannot restrict
our consideration to $\gamma$-exchange graphs only, because in this case
the squared transfer momentum $Q^2$ is so high that weak effects begin to
play an essential role in the total cross section and spin asymmetries.
At
the same time the ratio $m^2/Q^2$ (where $m$ is the mass of a scattering
lepton) becomes so small that it could be restricted to non-vanishing
terms for $m\rightarrow 0$.  The such calculation was already done by us
\cite{AISh}, but there we used only the naive parton model.

In this subsection the electroweak correction to the lepton
currents within QCD-improved parton distribution.

Generally, the double differential cross section lepton-nucleon
scattering in the frame electroweak theory has a form
\begin{equation}
\label{1.eq4}
\sigma _0= {4\pi \alpha ^{2} S_{x}S\over \lambda  _{s}Q^4}\Biggl[
L^{\gamma\gamma}_{\mu \nu}W_{\mu \nu}^{\gamma \gamma}(p,q)
+\frac 12(L^{\gamma Z}_{\mu \nu}
+L^{Z\gamma }_{\mu \nu})W_{\mu \nu}^{\gamma Z}(p,q)\chi
L^{ZZ}_{\mu \nu}W_{\mu \nu}^{ZZ}(p,q)\chi ^2
\Biggr ],
\end {equation}
where
$\chi=Q^2/(Q^2+M_Z^2)$ ($M_Z$ is the $Z$-boson mass), while the
explicit expression for leptonic
$L^{ij}_{\mu \nu}$
and
hadronic $W^{ij}_{\mu \nu}$ tensor
($i,j=\gamma ,Z$)
can be found in
\cite{LRC}.

After tensor contractions cross section (\ref{1.eq4}) can be
written in more simple form
\begin{equation}
\sigma _0={4\pi\alpha ^{2}y\over Q^4}
\sum^{8}_{i=1}
\theta^B_i{\cal F}_{i}
\label {b2}
\end {equation}
through linear combination eight generalized structure functions:
\begin{equation}
\begin{array}{ll}
\displaystyle
{\cal F}_i=R^{\gamma}_V{\bar F}^{\gamma }_i+
\chi R^{\gamma Z}_V{\bar F}^{\gamma Z}_i+
\chi^2 R^Z_V{\bar F}^Z_i\;\;&
\displaystyle
(i=1,2,6-8),\\[3mm]
\displaystyle
{\cal F}_i=R^{\gamma}_A{\bar F}^{\gamma }_i+
\chi R^{\gamma Z}_A{\bar F}^{\gamma Z}_i+
\chi^2 R^Z_A{\bar F}^Z_i&
\displaystyle
(i=3-5).
\end{array}
\label {calf}
\end{equation}

Here the quadratic combinations of the electroweak coupling
constants are defined as:
\begin{equation}
\begin{array}{l}
\displaystyle
 R^{mn}_V=(v^mv^n+a^ma^n)-P_L(v^ma^n+v^na^m),\;\;
\\[3mm]
\displaystyle
 R^{mn}_A=(v^ma^n+a^mv^n)-P_L(v^mv^n+a^na^m),
\end{array}
\label{1.eq10}
\end{equation}
where
\begin{equation}
\begin{array}{ll}
\displaystyle
v^{\gamma }=1,\qquad&
\displaystyle
v^Z=(-1+4s_w ^2)/4s_w c_w,
\\
a^{\gamma }=0,&
\displaystyle
a^Z=-1/4s_wc_w,
\end{array}
\label{1.eq6}
\end{equation}
$c_w$ and $s_w$ are cosin and sine of Weinberg's angle respectively
and $P_L$ is degree of lepton polarization.
The quantities $\theta^B_i$
depend only on the target polarization vector
and kinematic invariants:
\begin{equation}
\begin{array}{ll}
\displaystyle
\theta^B_1=Q^2,&
\displaystyle
\theta^B_5=\eta q Q^2S_p/2M^3,\\[0.2cm]
\displaystyle
\theta^B_2=(SX-M^2Q^2)/2M^2,\;\;&
\displaystyle
\theta^B_6=-(X\eta k_1+S\eta k_2)/2M,\\[0.2cm]
\displaystyle
\theta^B_3=Q^2S_p/4M^2,&
\displaystyle
\theta^B_7=\eta q(SX-M^2Q^2)/2M^3,\\[0.2cm]
\displaystyle
\theta^B_4=-Q^2\eta ( k_1+k_2)/M,&
\displaystyle
\theta^B_8=-\eta qQ^2/M.
\end{array}
\label{tb}
\end{equation}

Model independent part of lowest-order electroweak correction which
includes contributions both from a real photon emission
($\sigma_R$), and  from the additional virtual particles
($\sigma_V$) can be presented as a sum of infrared free terms:
\begin{equation}
\sigma _{RC}=
\sigma_V+\sigma _R=
\frac{\alpha}{\pi}\delta_{VR}\sigma _0+\sigma^r_V
+\sigma^F_R+{\hat\sigma} _R.
\label {all}
\end {equation}

The factor
\begin {eqnarray}
\delta _{VR}=&&
(\log \frac {Q^2}{m^2}-1)\ln
\frac{(W^2-(M+m_{\pi})^2)^2}{(X+Q^2)(S-Q^2)}
+\frac 32\log \frac {Q^2}{m^2}-2
\nonumber \\&&
-\frac 12\log^2\frac{X+Q^2}{S-Q^2}
+{\rm Li}_2
\frac{SX-Q^2M^2}{(X+Q^2)(S-Q^2)}
-\frac{\pi^2}6
\end {eqnarray}
appears in front of the Born cross section after
cancellation of infrared divergence by summing of
an infrared part separated from $\sigma _R$ and that
part of virtual contribution which arises
from the lepton vertex graphs including an additional virtual photon.

The contribution from electroweak rest loop correction
can be written in terms of the Born cross section with
the following replacement:
\begin {equation}
\sigma_V^r=\sigma^ B\left(
R^{mn}_{V,A}\rightarrow \delta R^{mn}_{V,A}
\right),
\end {equation}
and $\delta R^{mn}_{V,A}$ defined in \cite{LRC}.

The infrared free part of the cross section of the
radiative process can be presented as a sum of two terms
$\sigma _R^F $ and ${\hat\sigma_ R}$. The first term includes
unpolarized part of the cross section as well as that part of
cross section which appears from leading term $\xi _0$ of lepton
polarization vector
(\ref{vecpol}). It looks like eq. (\ref{eq25})
\begin{eqnarray}
\sigma _R^F =&&\alpha ^{3}y
 \int\limits^{\tau_{max}}_{\tau_{min}}d\tau
 \sum^{8}_{i=1}
\biggl\{ \theta_{i1}(\tau)
\int\limits^{R_{max}}_{0}{dR\over R}
\left[ {{\cal F} _{i}(R,\tau )\over (Q^2+R\tau)^2}-{{\cal F}
_{i}(0,0)\over Q^4}\right]\nonumber
\\
&&+ \sum^{k_{i}}_{j=2}
\theta_{ij}(\tau)\int\limits^{R_{max}}_{0}
dR {R^{j-2}\over (Q^2+R\tau)^{2}}{\cal F}_{i}(R,\tau)
\biggl\}.
\label{FR}
\end{eqnarray}
where the integration variables and their limits are defined by
formulae (\ref{rtau},\ref{rtau2}).
The second term containing the contribution which is proportional to
the rest part of polarization vector $\xi _1$
in the limit $m \rightarrow 0$  can be written through the Born
cross section:
\begin{equation}
 {\hat\sigma_ R}=\frac{\alpha y}{\pi S}\int\limits^{R^s_{max}}_0
\frac{RdR}{(S_x-R)}\tilde{\sigma}_{pl}^B.
\label{hat}
\end{equation}
The upper integration limit defines as
$R^s_{max}=S(W^2-(M+m_{\pi})^2)/(S-Q^2)$,
and $\tilde{\sigma}_{pl}^B$ is the proportional
$P_L$ part of the Born cross section with the following
replacement kinematic variable:
$S\rightarrow
S-R$, $Q^2\rightarrow Q^2(1-R/S)$ and $k_1\eta \rightarrow
k_1\eta (1-R/S)$.

The double differential cross section as a function of
the polarization characteristics of the scattering particles can be
presented as the sum of four terms:
\begin{equation}
\sigma=\sigma^u+P_L\sigma^{\xi}+P_N\sigma^{\eta}
+P_NP_L\sigma^{\xi\eta},
\label{cs}
\end{equation}
the first of them is an unpolarized cross section and three others
characterize the polarized contributions independent on polarization
degrees.There are no problems with the luminosity measurement in the
current collider experiments, so apart from the usual measurement of
polarized asymmetries the absolute measurement of the cross sections with
different polarization configurations of beam and target will be
probably possible in future polarization experiments at collider.
Besides, now the new methods of data processing, when experimental
information of spin observables is extracted directly from the
polarized part of the cross section \cite{Gagu,Gagu2} are actively
developed. In \cite{Gagu2} it is shown how to separate completely
unpolarized and polarized cross sections from a sample of
experimental data using a special likelihood procedure and a binnig
on polarization degrees. All above mentioned as well as the fact that
RC to asymmetry is always constructed from RC to parts of the cross
section allows to restrict our consideration to numerical studying of
RC to all of the cross sections in the equation (\ref{cs}) and their
combinations.

\begin{figure}
\vspace{10mm}
\unitlength 1mm
\begin{tabular}{cc}
\begin{picture}(60,60)
\put(-5,0){
\epsfxsize=6cm
\epsfysize=6cm
\epsfbox{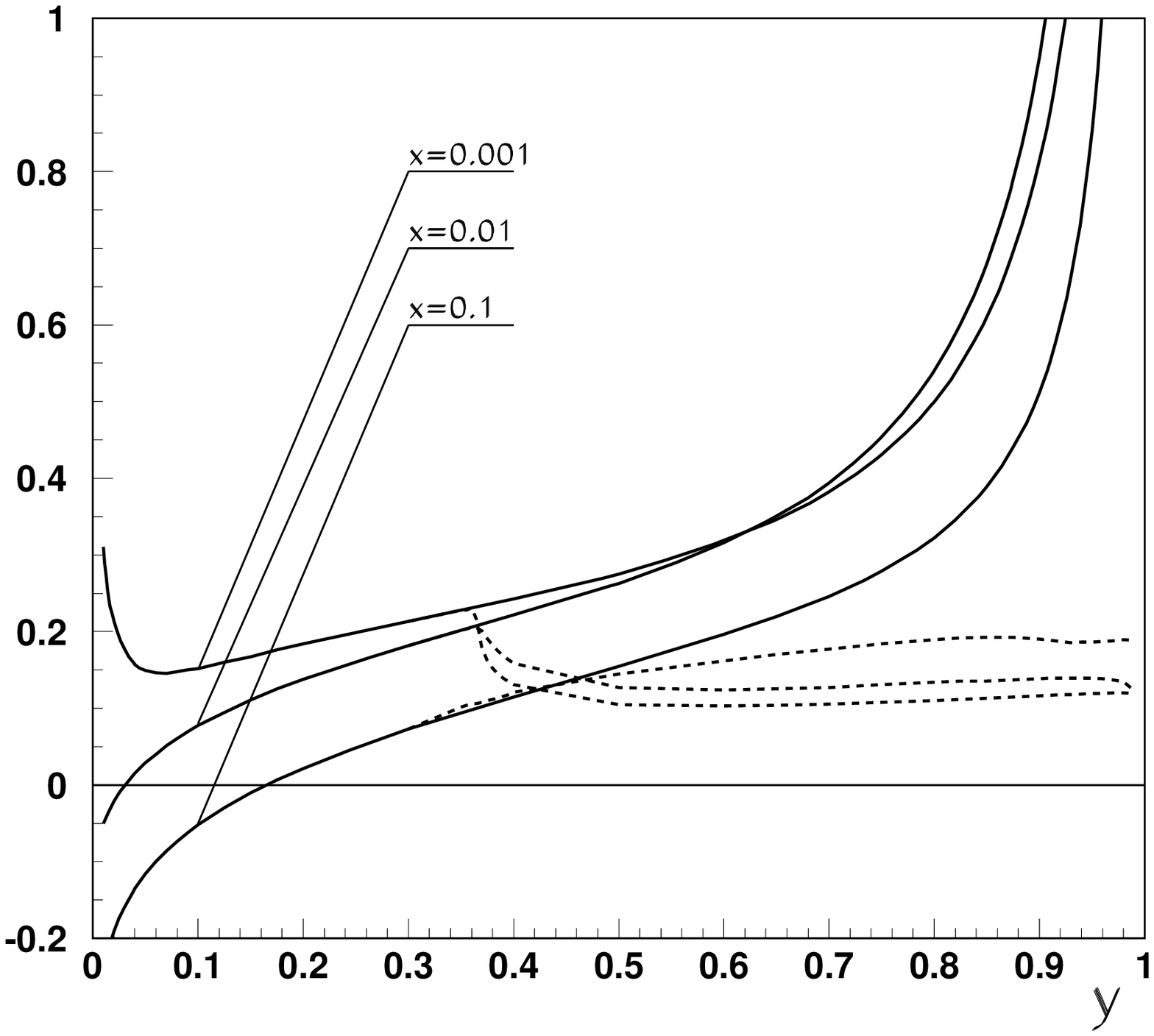}
}
\put(0,75){\makebox(0,0){$\delta^u$}}
\end{picture}
&
\begin{picture}(60,60)
\put(20,0){
\epsfxsize=6cm
\epsfysize=6cm
\epsfbox{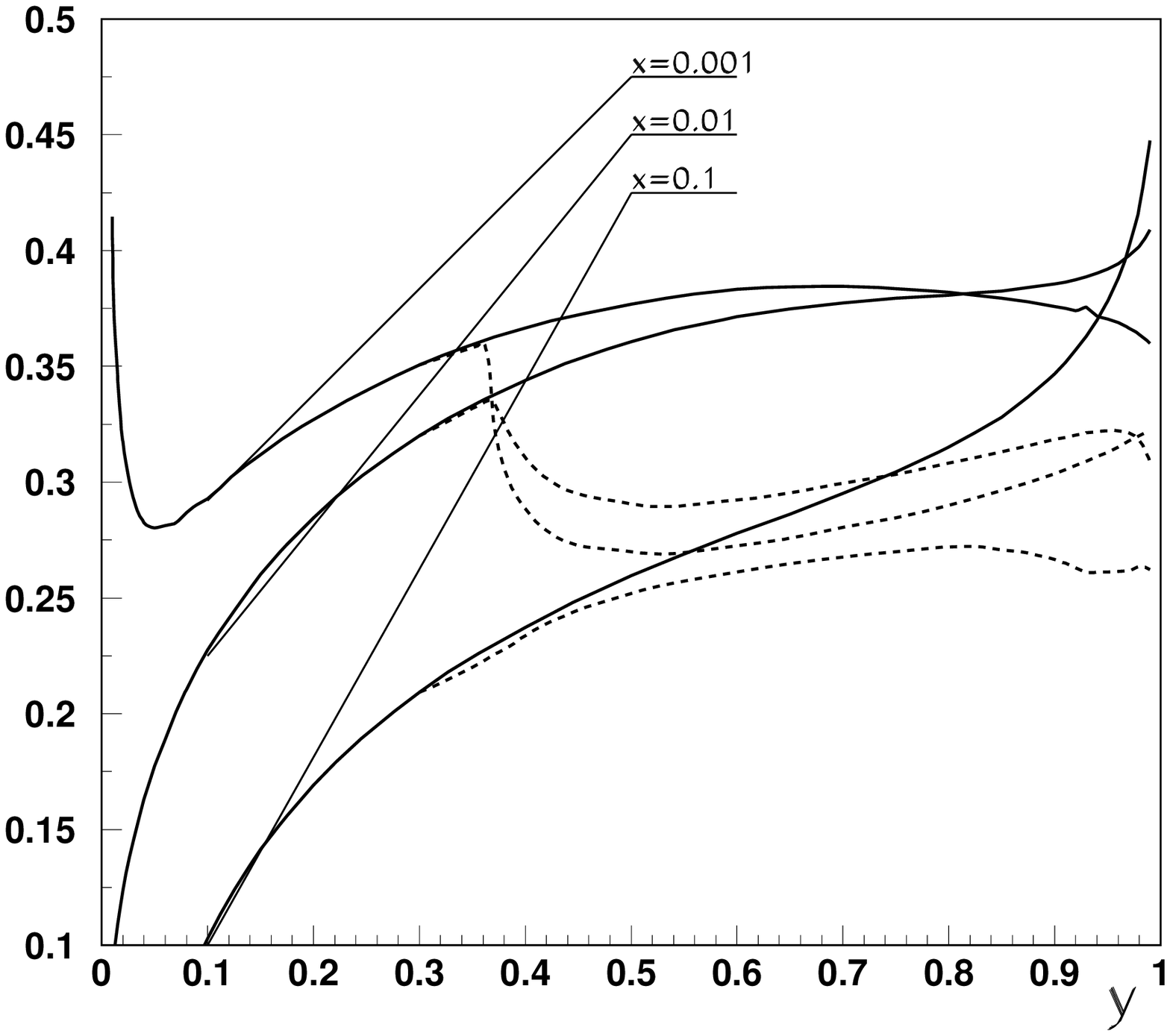}
}
\put(25,75){\makebox(0,0){$\delta^{\xi}$}}
\end{picture}
\vspace{2mm}
\\
\vspace{2mm}
\begin{picture}(60,60)
\put(-5,0){
\epsfxsize=6cm
\epsfysize=6cm
\epsfbox{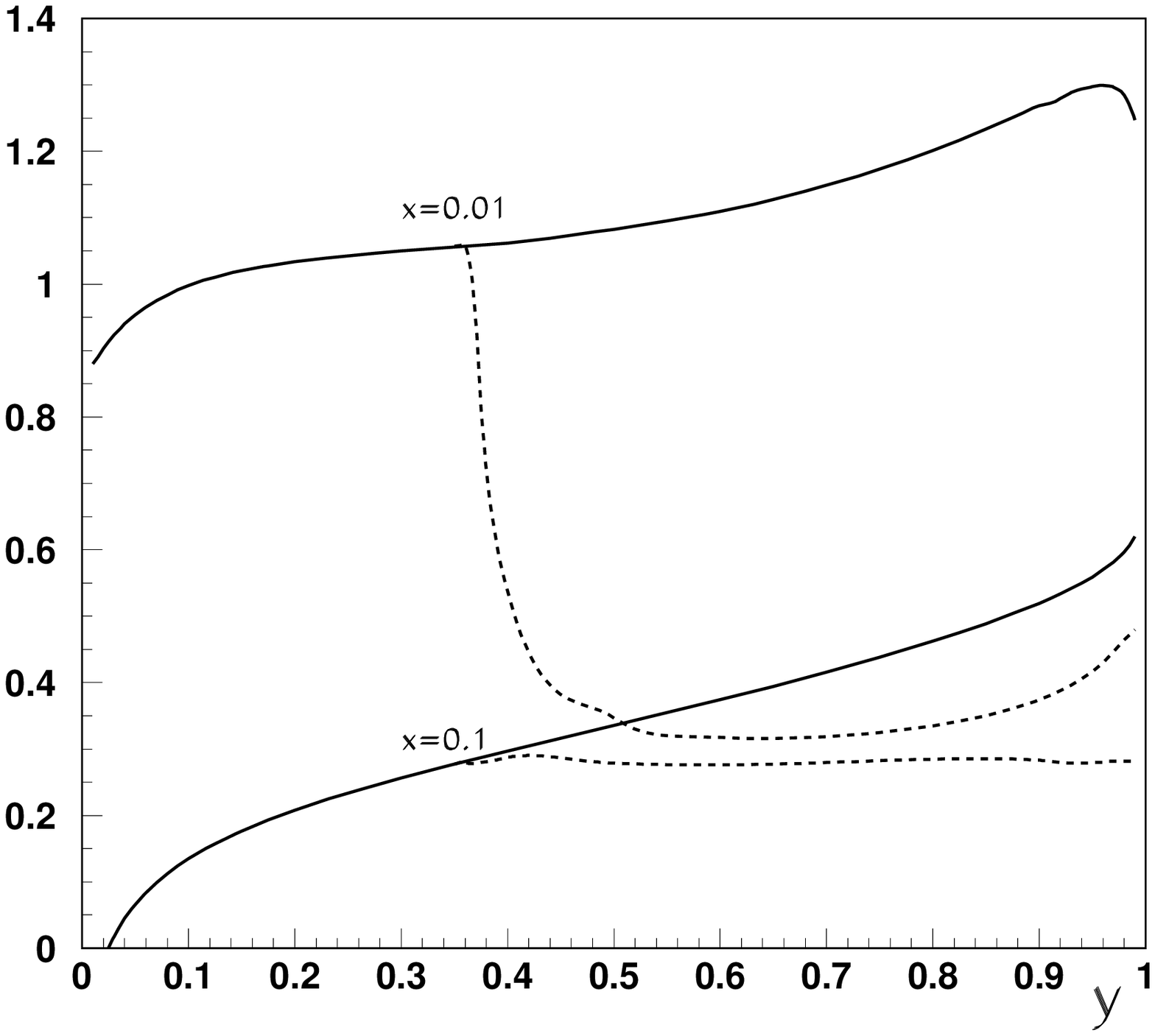}
}
\put(0,75){\makebox(0,0){$\delta^{\eta}_{||}$}}
\end{picture}
&
\begin{picture}(60,60)
\put(20,0){
\epsfxsize=6cm
\epsfysize=6cm
\epsfbox{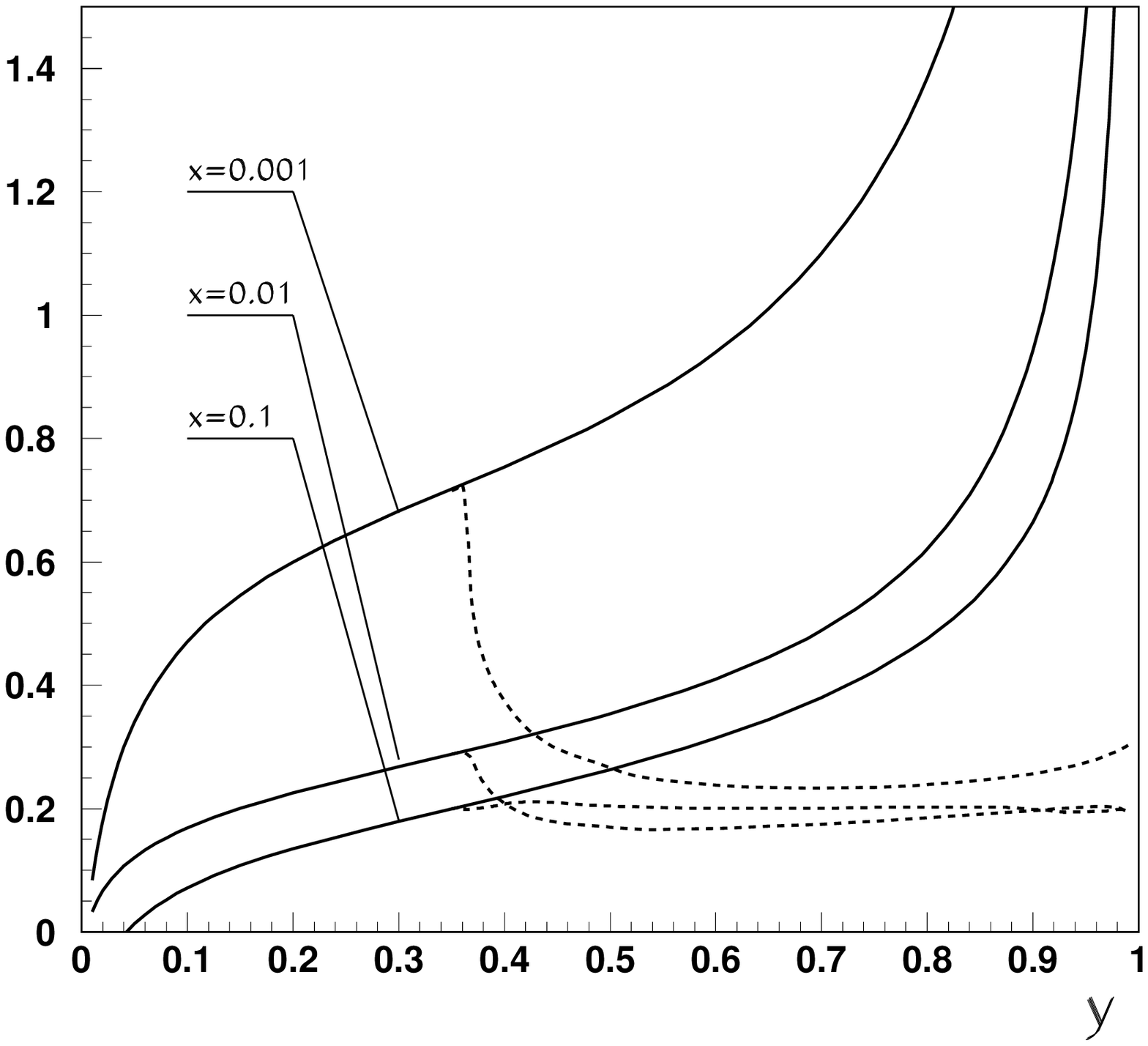}
}
\put(25,75){\makebox(0,0){$\delta^{\xi \eta }_{||}$}}
\end{picture}
\\
\begin{picture}(60,60)
\put(-5,0){
\epsfxsize=6cm
\epsfysize=6cm
\epsfbox{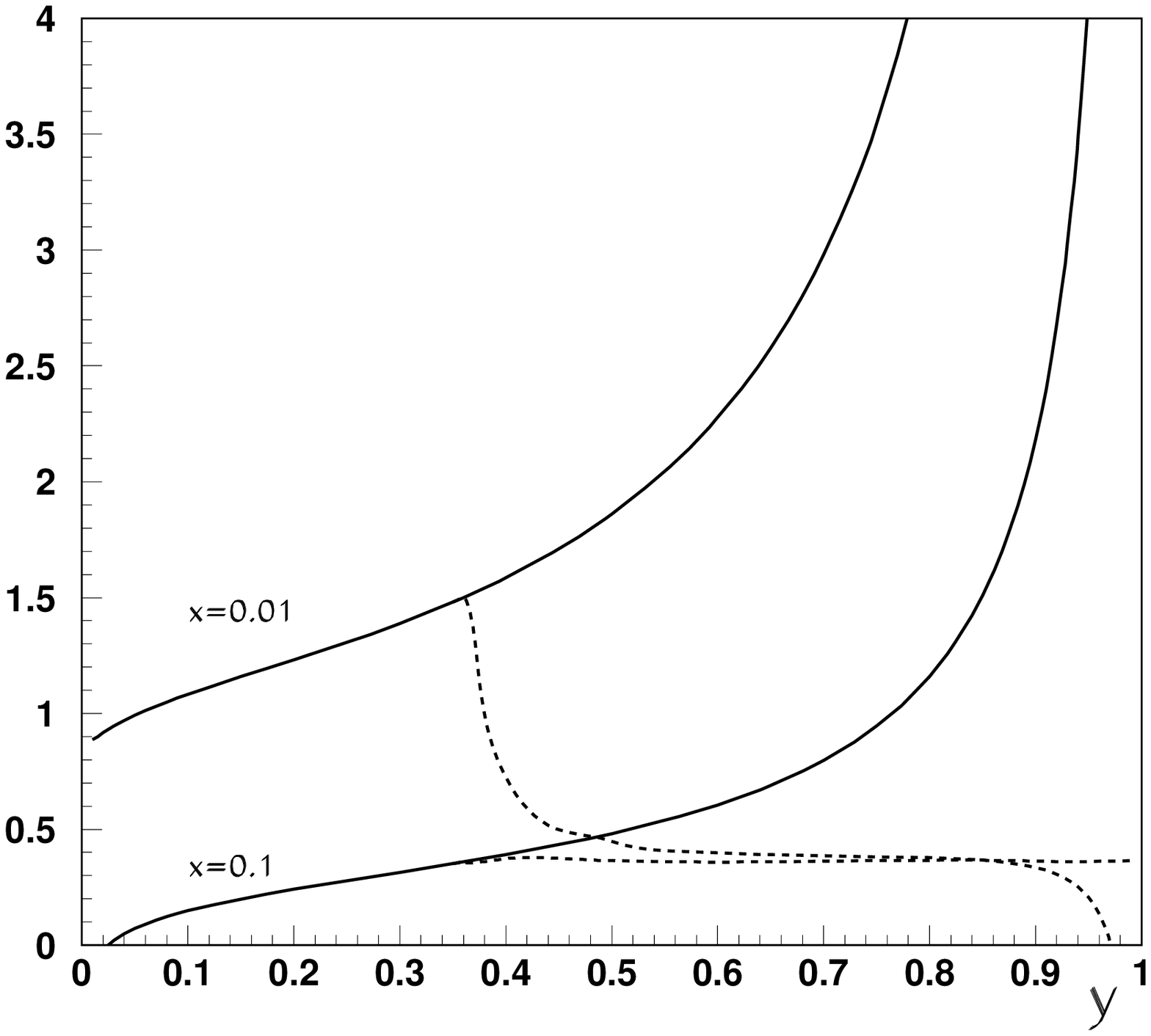}
}
\put(0,75){\makebox(0,0){$\delta^{\eta}_{\bot}$}}
\end{picture}
&
\begin{picture}(60,60)
\put(20,0){
\epsfxsize=6cm
\epsfysize=6cm
\epsfbox{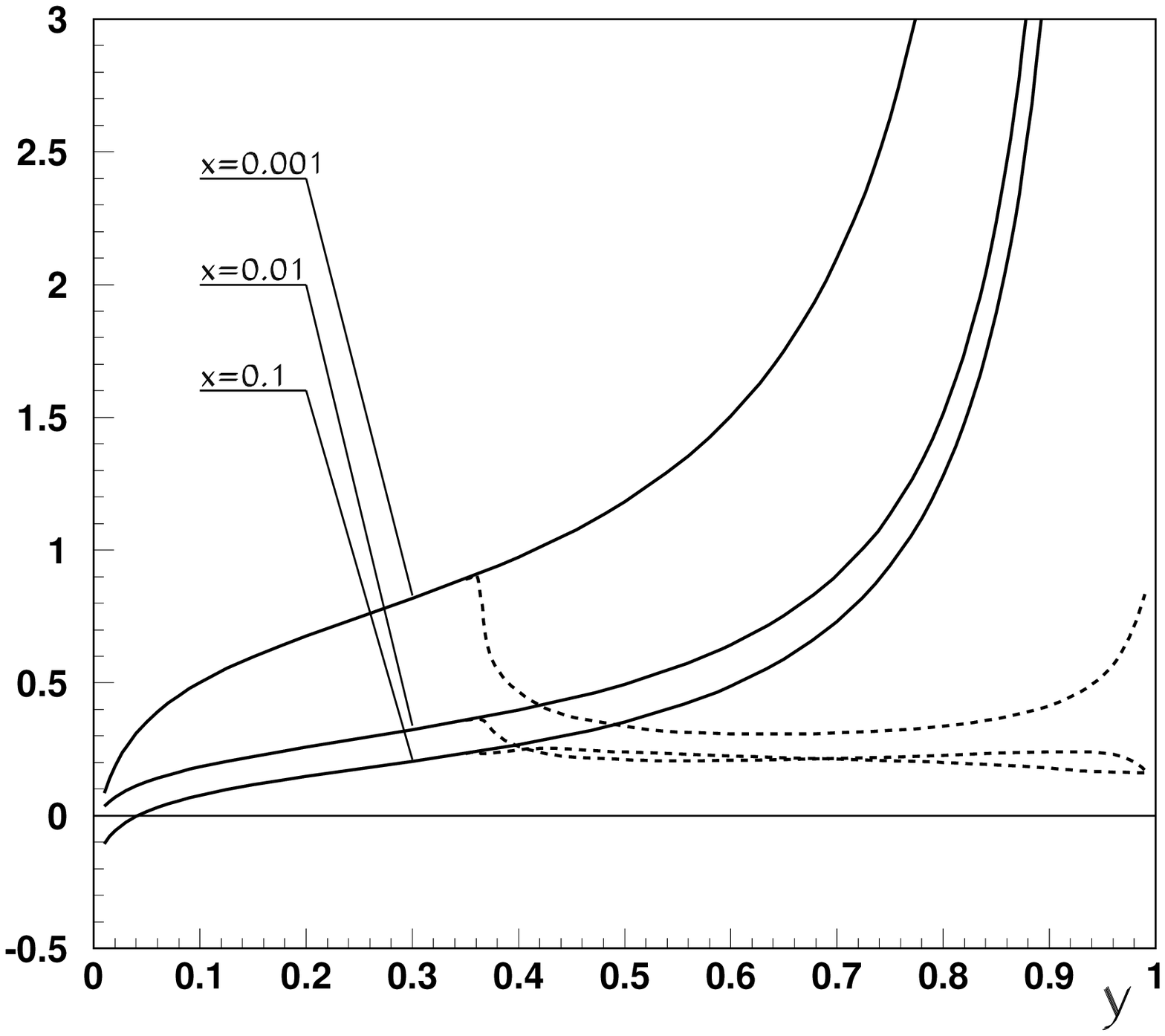}
}
\put(25,75){\makebox(0,0){$\delta^{\xi \eta}_{\bot}$}}
\end{picture}
\end{tabular}
\vspace{-15mm}
\caption{\protect\it
Radiative correction to
$\Delta \sigma$ defined in
(\ref{asyew}) with (dashed curves) and without cut (full curves).
 The down indexes $||$ and $\protect\bot $ correspond to
longitudinally
and transversely polarized proton beam respectively.
}
\label{Fg2}
\end{figure}

Radiative corrections to these cross sections defined as a ratio
of the cross section including one-loop RC only to the Born one
\begin{equation}
\delta^a=\frac{\sigma_{RC}^a}{\sigma_B^a}
=\frac{\sigma_{tot}^a}{\sigma_B^a}-1,
\qquad
(a=u, \xi , \eta , \xi \eta)
\label{asyew}
\end{equation}
is presented  on fig. \ref{Fg2}  as a function of
scaling variables $x$ and $y$. The kinematic region
corresponds to the  present unpolarization collider experiment at
HERA \cite{herau}.

 In experiments at collider a detection of a hard photon in calorimeter is
used to reduce radiative effects.  The dashed lines on fig. \ref{Fg2}
demonstrate the influence of an experimental cut on the RC. One of the
simplest variant of the cut when events having a radiative photon energy
$E_{\gamma } > 10$GeV are rejected from analysis, is considered only.

\section{Correction to the hadronic current in DIS}
\subsection{Tools of calculation and explicit results}

On the language of the hadronic tensor $W_{\mu \nu}$
the lowest-order one-loop
RC to the hadronic current consists of two parts
whose contributions are calculated in a different way:
$W^{1-loop}_{\mu \nu}=W^R_{\mu \nu}+W^V_{\mu \nu}.$
The first one appears from a gluon emission and requires the integration
over its phase space, while
the second part comes from a
gluon exchange graph and reads
\begin{equation}
W_{\mu \nu}^V=\frac 43 \frac {\alpha _s}{\pi}
\sum _q \biggl [-2\left ({\cal P}^{IR}+\ln \frac {m_q}{\mu} \right
)(l_q-1)
 -\frac 12 l_q^2
+\frac 32 l_q-2+\frac{\pi ^2}6\biggl ]
W_{\mu \nu}^{0q}+
W_{\mu \nu}^{AMM},
\end{equation}
where $W^{0q}_{\mu \nu}$ is a contribution of $q$-quark to  the hadronic
tensor on the Born level, and $W_{\mu \nu}^{AMM}$
is the quark anomalous magnetic moment.
The pole term which corresponds to the infrared divergence is contained in
${\cal P}^{IR}$. The arbitrary parameter $\mu $ has a dimension of a
mass.

 Both of these contributions include the infrared divergences, which have
to be careful considered in order to be canceled. Like QED we use the
identity
$$
W^R_{\mu \nu }=W^R_{\mu \nu }-W^{IR}_{\mu \nu }+W^{IR}_{\mu \nu }=
W^{F}_{\mu \nu }+W^{IR}_{\mu \nu }.
$$
Here $W^{F}_{\mu \nu }$ is finite for $k \rightarrow 0$, and
$W^{IR}_{\mu \nu }$ is
the infrared divergent part of $ W^R_{\mu \nu }$. Using the
dimensional
regularization scheme the latter can be given in the form
\begin{eqnarray}
W^{IR}_{\mu \nu }&=&\frac 4 3 \frac {\alpha }{\pi }\sum _q
\biggl[2\biggl ({\cal P}^{IR}+\ln \frac {m_q}{\mu} \biggr)(l_q-1)
+l_ql_v+\frac 12 l_q^2 \qquad
\nonumber \\ &&\qquad\qquad\qquad
-\frac 12 l_v^2
-\frac 34 l_q-\frac 74 l_v+\frac 34 -\frac{\pi ^2}3\biggr]
W^{0q}_{\mu
\nu },
\nonumber
\end{eqnarray}
where $ l_q=\log (Q^2/m_q^2)$, $l_v=\log ((1-x)/x)$.
The sum of $W_{\mu \nu}^{IR}$ and
 $W_{\mu \nu}^V$
\begin{eqnarray}
\label{delq}
W_{\mu \nu}^{IR}+
W_{\mu \nu}^V
&=&\frac  23 \frac {\alpha _s}{\pi}\sum _q
\bigl[2l_ql_v-l_v^2+\frac 32 l_q
-\frac 72 l_v-\frac 52
-\frac {\pi^2}{3}
\bigr]W_{\mu \nu}^{0q}
+W_{\mu \nu}^{AMM}
\nonumber \\ 
&=&\frac 23 \frac {\alpha _s}{\pi}\sum _q\delta _qW_{\mu \nu}^{0q}
+W_{\mu \nu}^{AMM}
\end{eqnarray}
is infrared free.

In order to extract some information about QCD contribution to the
polarized structure functions, the integration in $W^F_{\mu \nu}$
over the gluon phase space should be performed without any assumptions
about the polarization vector $\eta $. So the technique of tensor
integration have to be applied in this case. Since the result of the
analytical integration has the same tensor structure as the usual hadronic
tensor in polarized DIS, the coefficients in front of the corresponding
tensor structures (like $g_{\mu\nu}$, $p_\mu p_\nu$ ... ) can be
interpreted as one-loop QCD contributions to the corresponding structure
functions.

Thus the QCD-improved structure functions $F_{1,2,L}$ and
$g_1$ read
\begin{eqnarray}
\label{qcdsf1}
F_1(x,Q^2)=
\frac 1{2x}[F_2(x,Q^2)-F_L(x,Q^2)],
&&
F_2(x,Q^2)=
x \sum _q e_q^2f_q(x,Q^2),
\nonumber \\
F_L(x, Q^2)=
\frac {4\alpha _s}{3\pi}x\sum _q e_q^2\int
\limits_x^1dzf_q(x/z),
&&
g_1(x,Q^2)=
\frac 12 \sum _q e_q^2\Delta f_q(x,Q^2),
\end {eqnarray}
and $g_2$ looks like
\begin{eqnarray}
\label{qcdsf2}
g_2(x,Q^2)=
\frac {\alpha _s}{6\pi}\sum _q e_q^2
\biggl\{
(1-2l_q-\ln (1-x)
)\Delta f _q(x)
+\int \limits ^1 _xdz
\Bigl [ (4l_q
\nonumber \\
-4\log z(1-z)
-12
-\frac 1{(1-z)})\Delta f_q(x/z)
+\frac{\Delta f_q(x)}{(1-z)}
\Bigl ]
\biggl\},
\end {eqnarray}
where the $Q^2$-dependent unpolarized and polarized parton distributions
are defined as
\begin{eqnarray}
f_q(x, Q^2)=
(1+\frac {2\alpha _s}{3\pi} \delta _q)
f_q(x)
+\frac {2\alpha _s}{3\pi} \int\limits_x^1 \frac{dz}{z}
\Bigl[
(\frac {1+z^2}{1-z}
( l_q
-\log  z(1-z) )
\nonumber \\ \qquad \qquad
-\frac 72 \frac 1{1-z}
+3z
+4)f_q(\frac{x}{z})
-\frac 2{1-z}\left(l_q+\log \frac{z}{1-z}
-\frac 74\right)f_q(x)
\Bigl],
\nonumber \\
\Delta f_q(x,Q^2)=
(1+\frac {2\alpha _s}{3\pi} \delta _q)
\Delta f_q(x)
+ \frac {2\alpha _s}{3\pi} \int\limits_x^1 \frac{dz}{z}
\Bigl[(\frac{1+z^2}{1-z}
(l_q
-\log  z(1-z) )
\nonumber \\  \qquad \qquad
-\frac 72 \frac 1{1-z}
+4z
+1)\Delta f _q(\frac{x}{z})
-\frac 2{1-z}\left(l_q+\log \frac{z}{1-z}
-\frac 74\right)\Delta f _q(x)
\Bigl],
\label{qcdpd}
\end{eqnarray}
and $\delta _q $ can be found in (\ref{delq}).

It has to be noted that our formulae
(\ref{qcdsf1},\ref{qcdsf2},\ref{qcdpd}) are in the
agreement with  ones obtained earlier. The unpolarized structure
functions coincide with (2.24,2.49) from \cite{NE1}, $g_1(x,Q^2)$
corresponds to the expression (13) in \cite{NE2}. At last
$g_2(x,Q^2)$
can be considered with the sum (18) and (19) from \cite{AG2}.

After integration of the QCD-improved structure functions over the
scaling variable $x$ it could be seen that QCD RC to the first
moment of the unpolarized structure functions as well as $g_2$
have the identical values both for massive and massless quark
approaches. However the
value of QCD-correction to the first moment of $g_1$
depends on the scheme. It was calculated:
\begin{equation}
\int \limits _0^1g_1(x,Q^2)=
(1-C_{g1}\frac {\alpha_s}{\pi})
\int \limits _0^1 dx g_1^0(x),
\label{sr}
\end{equation}
where  $C_{g1}=1$ for massless and $C_{g1}=5/3$  for massive one.

\subsection{Discussion and Conclusion}

In this section we applied the approach traditionally used in
QED
and electroweak theory for calculation of QCD correction to the
DIS structure
functions and sum rules. Since the quark was considered massive
within this ap\-p\-ro\-ach, it allowed us to estimate the finite
quark mass effects
at the NLO level. LO correction  contributes to the DIS structure
functions but vanishes for the
sum rules, so this mass effects
are important just for the sum rules.

We found that there is no any additional effect for the first moment of
the unpolarized structure functions as well as for $g_2$. However there is
some non-zero correction to the first moment of $g_1$ and as a result
to the
Ellis-Jaffe and Bjorken sum rules. The value of the correction is in
agreement with the results of refs.\cite{NE2,TERV}, obtained by different
methods. We confirm also the statement of \cite{NE2} that the
classical
value of correction to the first moment of $g_1$ is reproduced if
we take into account the
leading term of the polarization vector $\eta=p_{1q}/m_q$. However
contrary to the paper we would interpret the result with $C_{g1}=5/3$
in eqs.(\ref{sr}) as
physical one. In our calculation we took the proton (quark)
polarization vector in a general form. Using of the exact
representation of the
vector \cite{ASh}: 
\begin{equation}
\eta=\frac1{\sqrt{(k_1p_{1q})^2-m^2m_q^2}}
\Bigl[\frac {k_1p_{1q}}{m_q}p_{1q}-m_qk_1\Bigr]
\label{et}
\end{equation}
leads to exactly the same result (with $C_{g1}=5/3$) for the first
moment of $g_1$. The second term giving the additional
contribution $\sim 2/3$ cannot be neglected within the
approximation
under consideration. It can be easy understood from pure
calculation of QED and electroweak corrections to the lepton current
in DIS process \cite{BARD}, where the lepton polarization vector
looks like hadron one (\ref{vecpol}).
The explicit expression of this non-vanishing for $m \rightarrow 0$
contribution within electroweak correction to the lepton current is
defined by the formula (\ref{hat}) from previous section. The more
detailed description of this contribution treatment can be found
in \cite{LRC}.

The result with $C_{g1}=5/3$ was
obtained under similar assumptions in the report \cite{TER}. The
additional correction was discussed both within OPE (operator
product expansion) and within improved quark-parton model in
paper \cite{TERV}. It was shown in this paper that there is no
contradiction between this result and classical correction obtained for
massless QCD, if we carefully take into account the finite
quark mass effects for coefficient function and matrix element.
As it was reviewed in recent paper \cite{NE3} the
renormalization of the axial-vector current for massless quark completely
suppresses the additional contribution obtained within massive approach
(see section 4 of the paper).

We note that Burkhard-Cottingham sum rule is held within taking
into account mass effects at the NLO level. The target mass
correction of the next order ($\sim m_q^2/Q^2$) was analyzed and
was shown negative in the paper \cite{TERM}. However, in the case
of conserved currents (see \cite{BT}) the polarized structure
function $g_2$ with the target mass correction obeys not only
Burkhard-Cottingham sum rule but and Wandzura-Wilczek relation
too.

Estimating the diagrams with one-gluon radiation we have also the result
for QED radiative correction to had\-ro\-nic current. The calculation
within the quark parton model shows that QED correction is not so large,
however there are at least two arguments to be taken into account
it. First, the DIS structure functions are defined for the one photon
exchange approximation as an objects including only the strong
interaction. It means that transferring from the cross section to the
structure functions we neglect these QED effects, so their contributions
have to be included to systematic error of corresponding measurements.
Second reason is that there exist some measurements where the QED
correction is important. One of the example can be calculation of
$\alpha_s $ by comparing theoretical and experimental values for
the Bjorken sum rule \cite{EK}. In this case $\alpha_s$ is
suggested to be extracted from QCD corrections to the Bjorken sum
rule, which is a series over $\alpha_s $, and at least five terms
of the expansion are known. Simple estimation shows that the QED
contribution is comparable with already fourth (or even third)
term of the expansion.

There are several papers devoted to calculation of QED and electroweak
corrections to the hadronic current \cite{t-79,ShT,ShZ,AISh,BARD,SP}. As usual methods
similar ours are used for that. Positive moment here is the keeping quark
mass non-zero. From the other side there are some effects which are not
considered within the QED calculations: taking into account the confiment
effect in integration of soft region, consideration of the quarks as
non-free particles. Thus the methods (see for example \cite{TER})
developed for careful treatment of the QCD effects should be
applied for the analysis of photon emission from the hadronic
current. However, probably the best way to take into account the
photon
radiative correction to the hadronic current is further
generalization of OPE technique for including of the QED effects.

\section{Corrections to the lepton current in other processes}
\subsection{Diffractive vector meson electroproduction and
code DIFFRAD}
The measurement of the cross section of the exclusive vector meson
electroproduction
\begin{equation}\label{DF1}
 e(k_1) + p (p) \longrightarrow e'(k_2) + \vec {v}(p_h)+p (p_2),
\end{equation}
can provide information on the had\-ronic
component of the photon and on nature of diffraction.
During several
years the diffractive production of the vector meson has been the
subject of the muonproduction \cite{EMC,NMC,E665} and
electroproduction
\cite{H1,ZEUS,HERMES2} experiments.
Data analysis of these experiments is considerably affected by the QED
radiative effects.
At practice RC to the processes of
electroproduction are taken into
account
using codes originally developed for the inclusive case (see
ref.\cite{Kurek}, for example).

 The purpose of this subsection is to present the electromagnetic
correction to experimentally observed cross sections that was
calculate by Bardin-Shumeiko approach in ref.\cite{DIFF}.

We consider RC
to three and four dimensional cross sections
${\sigma}= d\sigma/dxdydtd\phi_h$ and
${\bar\sigma}= d\sigma/dxdydt$. They are related as
\begin{equation}
{\bar\sigma}=\int\limits_0^{2\pi} d\phi_h \; \sigma \; .
\label{eq0}
\end{equation}
 The four differential Born cross section
can be presented
 in the form
\begin{equation}
\sigma_0 =
\frac{\alpha}{4\pi^2xy}
\biggl(y^2\sigma_T+2\bigl(1-y-\frac{1}{4}y^2\gamma^2\bigr)
(\sigma_L+\sigma_T)\biggr),
\label{born}\end{equation}
where $x$ and $y$ are usual scaling variable,
$\sigma_T$ and $\sigma_L$ are differential cross sections of
the photoproduction, $\gamma^2=Q^2/\nu^2$ and $\nu$ is the virtual photon
energy.

For the observed cross section of the vector meson electroproduction
we obtain
\begin {equation}
\sigma _{obs} = \sigma _0 e^{\delta_{inf}}
(1 + \delta_{VR}+\delta_{vac})+
\sigma_{F}.
\label{eq134}
\end {equation}

As earlier the correction $\delta_{vac}$  comes from the
effects of vacuum polarization by leptons and hadrons.

The contribution of the infrared finite part can be written
in terms of POLRAD 2.0 notation \cite{ASh,P20}:
\begin{eqnarray}\label{eq253}
\sigma _F = -{\alpha^2 y\over 16 \pi^3}
 \int\limits^{2\pi }_{0}d\phi_k
 \int\limits^{\tau_{max}}_{\tau_{min}}d\tau
\sum_{i=1}^2\sum_{j=1}^3 \theta_{ij}
\int\limits^{v_{m}}_{0}
{dv\over f}
R^{j-2}
\biggl[
{{\cal F}_i \over {\tilde Q}^4 }
-\delta_j{{\cal F}_i^0 \over Q^4 }
\biggr]
,
\end{eqnarray}
where $R=v/f$, $f=1+\tau-kp_h/kp$ and
$2M\tau_{max,min}=S_x\pm \sqrt{\lambda_q}$; as usual $\delta_j=1$
for
$j=1$ and  $\delta_j=0$ otherwise.
The quantities $\theta_{ij}$ depend only on the kinematic invariants
and the integration variables $\tau$ and $\phi_k$. The explicit
expressions for them can be found in \cite{DIFF}.
$\theta $-functions are kinematical ones without any integrals
for
unpolarized case, and only one integration are left in the case of
polarization part.

The dependence on the photoproduction cross sections is included in ${\cal
F}_i$:
\begin{equation}
\begin{array}{ll}
\displaystyle
{\cal F}_1=(S_x-R){\sigma}_{T}^R, &
\displaystyle
\;\;{\cal F}_2={2{\tilde Q}^2\over
S_x-R}({\sigma}_{T}^R+{\sigma}_{L}^R),
\\[0.3cm]
\displaystyle
{\cal F}_1^0=S_x\sigma_T, &
\displaystyle
\;\;{\cal F}_2=2x(\sigma_T+\sigma_L).
\end{array}
\end{equation}
The quantities ${\sigma}_{T,L} $ have to be calculated for Born
kinematics, but ${\sigma}_{T,L}^R$ is calculated in terms of
so called true kinematics. It means that they have to be calculated
for the tilde variables
instead of standard $Q^2$, $W^2$ and $t$ 
which have being already defined
by
eq. (\ref{til}).

The important point is  the dependence of the results on the
maximal inelasticity
$v_{m}$. The
inelasticity is calculated in terms of the measured momenta, so it is
possible
to make a cut on the maximal value of this quantity.  If this cut is not
applied the maximal inelasticity is defined by kinematics only. Below we
give the formulae for
$ v_{m}$ in terms of the kinematic invariants
\begin{eqnarray}\label{vmkin}
4Q^2v_{m}=\Biggl(\sqrt{\lambda_q}-\sqrt{t_q^2+4m_v^2Q^2}\Biggr)^2-
(S_x-2Q^2+t_q)^2-4M^2Q^2.
\end{eqnarray}

The
FORTRAN code DIFFRAD created on the basis of the presented
formulae
 calculates the lowest order RC to the diffractive vector meson
electroproduction. The higher order effects are approximated by the
procedure of exponentiation. The formulae  for the cross section are given
in a covariant form, so the code can be run both for the fixed target
experiments and for the experiments at collider.

Below we give numerical results for RC factor
\begin{equation}
\eta={\bar\sigma_{obs}\over\bar\sigma_0}
={\int\limits_0^{2\pi}d\phi_h\sigma_{obs}\over
\int\limits_0^{2\pi}d\phi_h\sigma_0}
\end{equation}
obtained
within the kinematic regions of recent HERMES and
HERA experiments for exclusive $\rho(770)$ electroproduction.

\begin{figure}[t]\centering
\unitlength 1mm
\hspace*{-5mm}
\parbox{.4\textwidth}{\centering
\begin{picture}(80,80)
\put(62,8){\makebox(0,0){$Q^2$,GeV$^2$}}
\put(8,77){\makebox(0,0){ $\eta$}}
\put(-5,5){
\epsfxsize=8cm
\epsfysize=8cm
\epsfbox{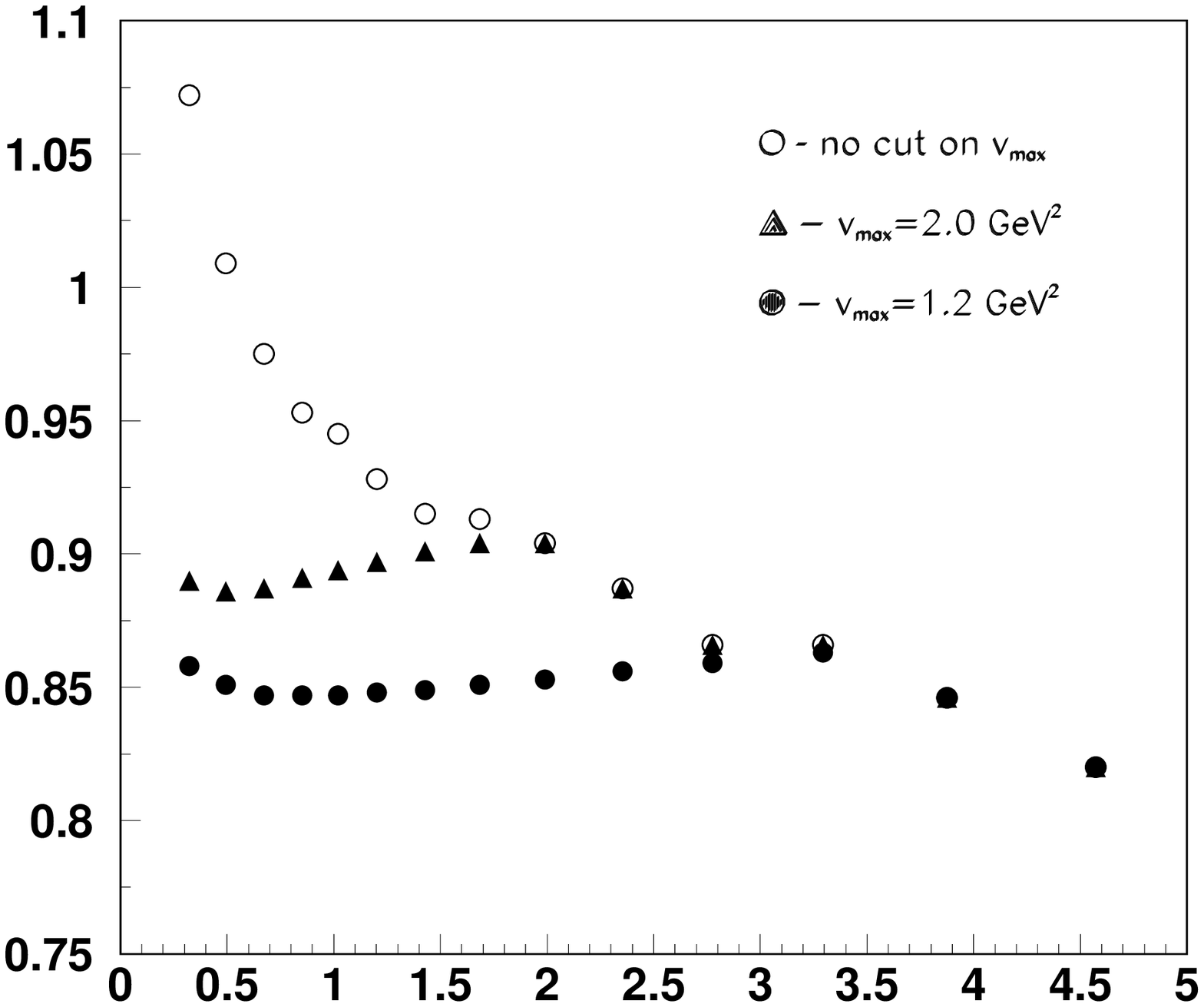}
}
\end{picture}
\vspace*{-10mm}
\caption{\protect \it RC factor under kinematic conditions of HERMES;
$\sqrt{S}=7.9$ GeV, $t=-0.11$ GeV$^2$.}
}
\label{f43}
\hspace*{8mm}
\vspace*{10mm}
\parbox{.4\textwidth}{\centering
\begin{picture}(80,80)
\put(62,5){\makebox(0,0){\small  $-t$,GeV$^2$}}
\put(52,40){\makebox(0,0){\bf observed}}
\put(42,25){\makebox(0,0){\bf Born}}
\put(12,72){\makebox(0,0){\small  $\sigma, mb$}}
\put(-8,0){
\epsfxsize=8cm
\epsfysize=8cm
\epsfbox{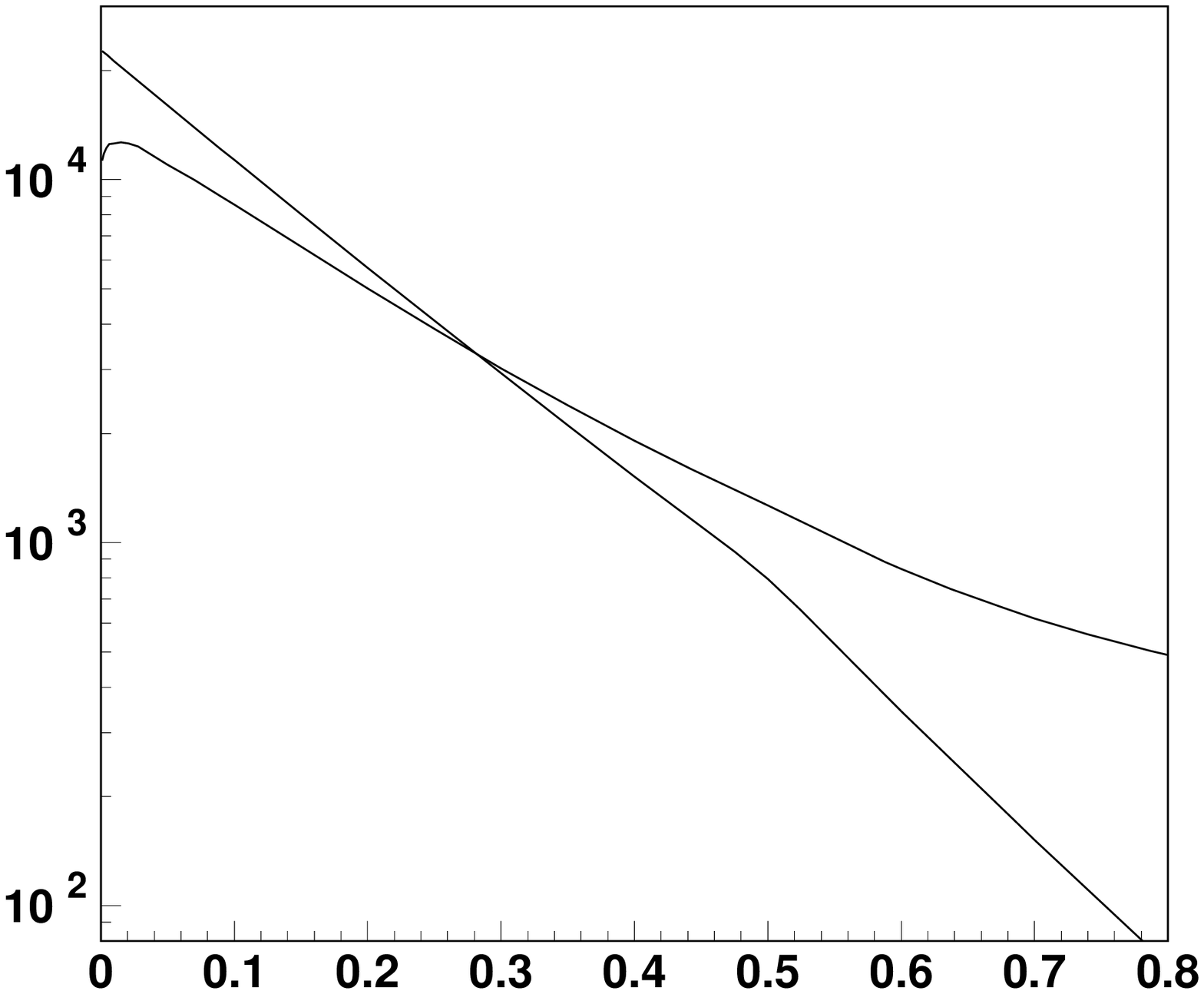}
}
\end{picture}
\caption{
\label{tdep}
\protect \it
Observable and Born cross sections
as a function of $t$ under kinematic conditions of HERA;
$\sqrt{s}$=300
 GeV; $W$=70 GeV;
$Q^2$=4
 GeV$^2$.
}
}
\vspace*{-10mm}
\end{figure}

The dependence on the $v_{m}$ cut in the region of HERMES kinematics is
analyzed
in fig.8. The usage of the cut changes RC factor  for small
$Q^2$ and does not influence on  RC for  larger values of
$Q^2$.
In the case of the cut usage
we
have to define $v_m$ as a minimum of the value of the cut and $v_m$ given
by the kinematic restrictions (\ref{vmkin}).
For fixed values of $-t$ the cut influence on $v_m$ and RC till
certain value of $Q^2$ only.

The $Q^2$-dependence shown in this figure is typical
for the situation when the inelasticity cut is not applied. If this cut
is
used then the rise of $\eta$ when $Q^2$ goes down would be
suppressed. Notice that the $t$-dependence is
rather important too. Fig.\ref{tdep} illustrates this
last property. $\eta$ crosses unity for $-t \sim $ 0.25--0.3
GeV$^2$ and rises with increasing $|t|$. The large positive correction
in this case is a result of the large phase space for photon radiation.
The cut applied to the inelasticity (or $E-p_z$) can again reduce the
value of the RC factor for large $|t|$. As a consequence of the
$t$-dependence of $\eta$, the observed slope parameter also receives
large RC.

Now as application of the obtained result let us consider the
process
\begin{equation}
e(k_1) + p(p) \rightarrow e(k_2) + p(p') + \rho (p_V), \;\;\;
\rho \rightarrow \pi ^+ (p_+) + \pi ^- (p_-),
\label{eq1m}
\end{equation}
that can be viewed as an
off-diagonal Compton scattering analytically continued in the
virtuality of the photon $\gamma ^*$ to the vector meson mass
$\gamma ^* p \rightarrow V p$ and
gives access to the whole set of the corresponding helicity
amplitudes.

The process (\ref{eq1m}) is analyzed experimentally by
means of
spin-density matrix  elements. When measured, they
give an indication of vector meson internal constituents motion
and its spin-angular structure.
The angular distribution of unpolarized vector meson decay is
parameterized by fifteen matrix elements
$r_{ij}^{\alpha }, r_{ij}^{\alpha \beta }$.
For a long time it was believed, that their
behavior complies with
the s-channel helicity conservation (SCHC)  hypothesis, which means
that the helicity of the virtual photon is conserved in the
s-channel process $\gamma ^* p \rightarrow \rho p$. In this case
ten matrix elements (which corresponds to the
case when
photon and vector meson have different helicities) are equal to
zero. But in the recent measurements
$r^5_{00}$ has been observed to be non zero \cite{ZEUS1,H1a,ZEUS2}, what
has been considered as
an indication to SCHC violation.

The procedure of the experimental data analysis is based on the
correlation of the lepton scattering, vector meson production
and decay planes, which are affected by RC. Hence it is topical
to look at whether the measured $r^5_{00}$ can, at least partly,
be the result that RC coming from non-observed QED effects and real
photon emission was underestimated. In any case in order to make
the data processing of the corresponding experiments
\cite{ZEUS1,H1a,ZEUS2} to  be  consistent, RC should be taken into
account.

As  it was shown in \cite{AKuzh} the value of
the QED corrections ($\Delta r=r_{obs}-r_{Born}$)
to  the matrix elements are defined
by the following quantity:
\begin{equation}
I_{1}=\frac {\int\limits _0^{2\pi } d\phi _h\;
\cos \phi _h
\; \delta
(\phi _h )}
{\int\limits _0^{2\pi } d\phi _h\; (1+
\delta (\phi _h ))}, \;\;\; n=1,...,4,
\end{equation}
where $\delta (\phi _h )=\sigma_{obs}/\sigma _0-1$.

\begin{figure}
\unitlength 1mm
\vspace{-2cm}
\begin{center}
\begin{picture}(160,80)
\put(0,-10){
\epsfxsize=8cm
\epsfysize=8cm
\epsfbox{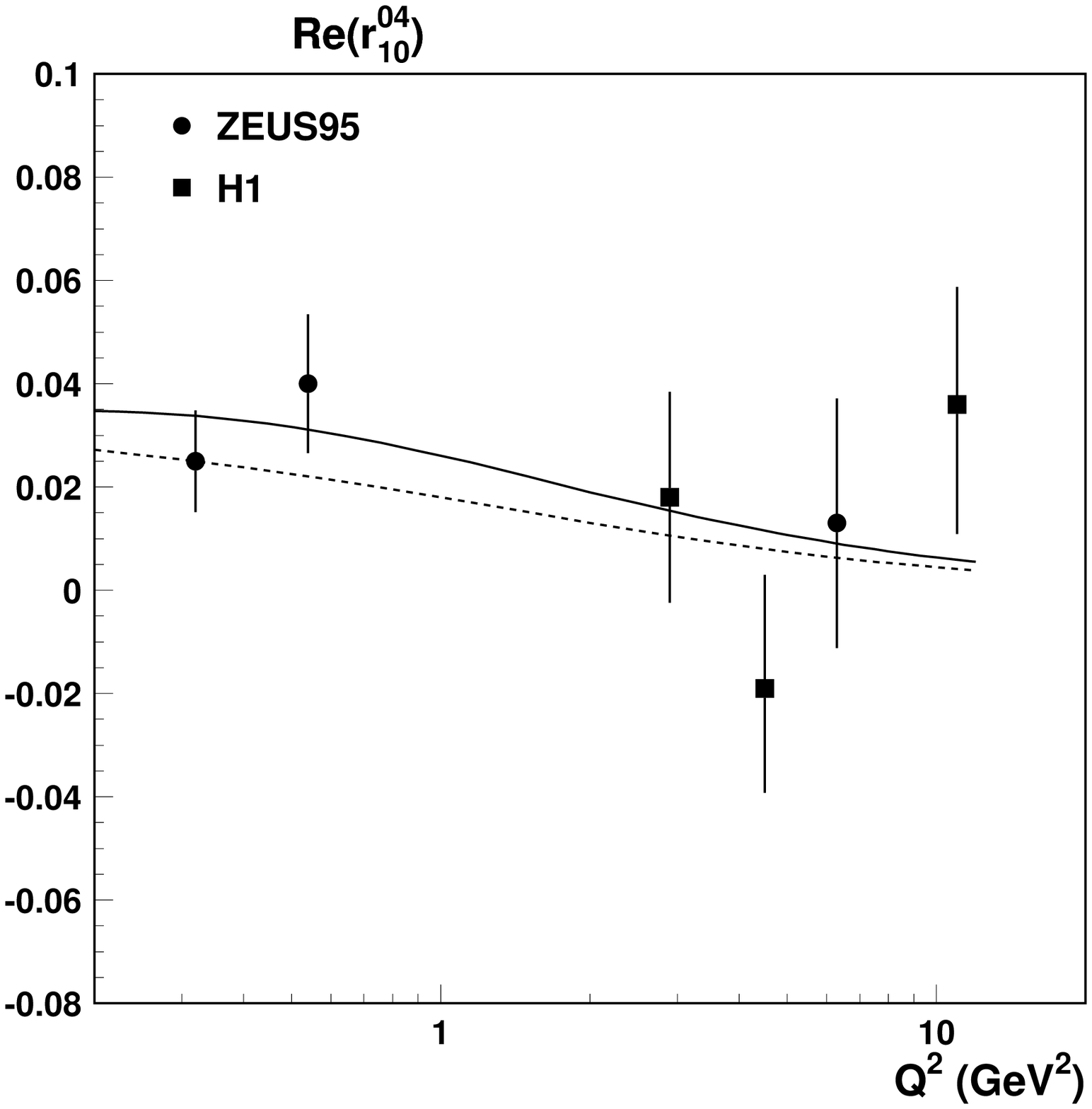}
}
\put(80,-10){
\epsfxsize=8cm
\epsfysize=8cm
\epsfbox{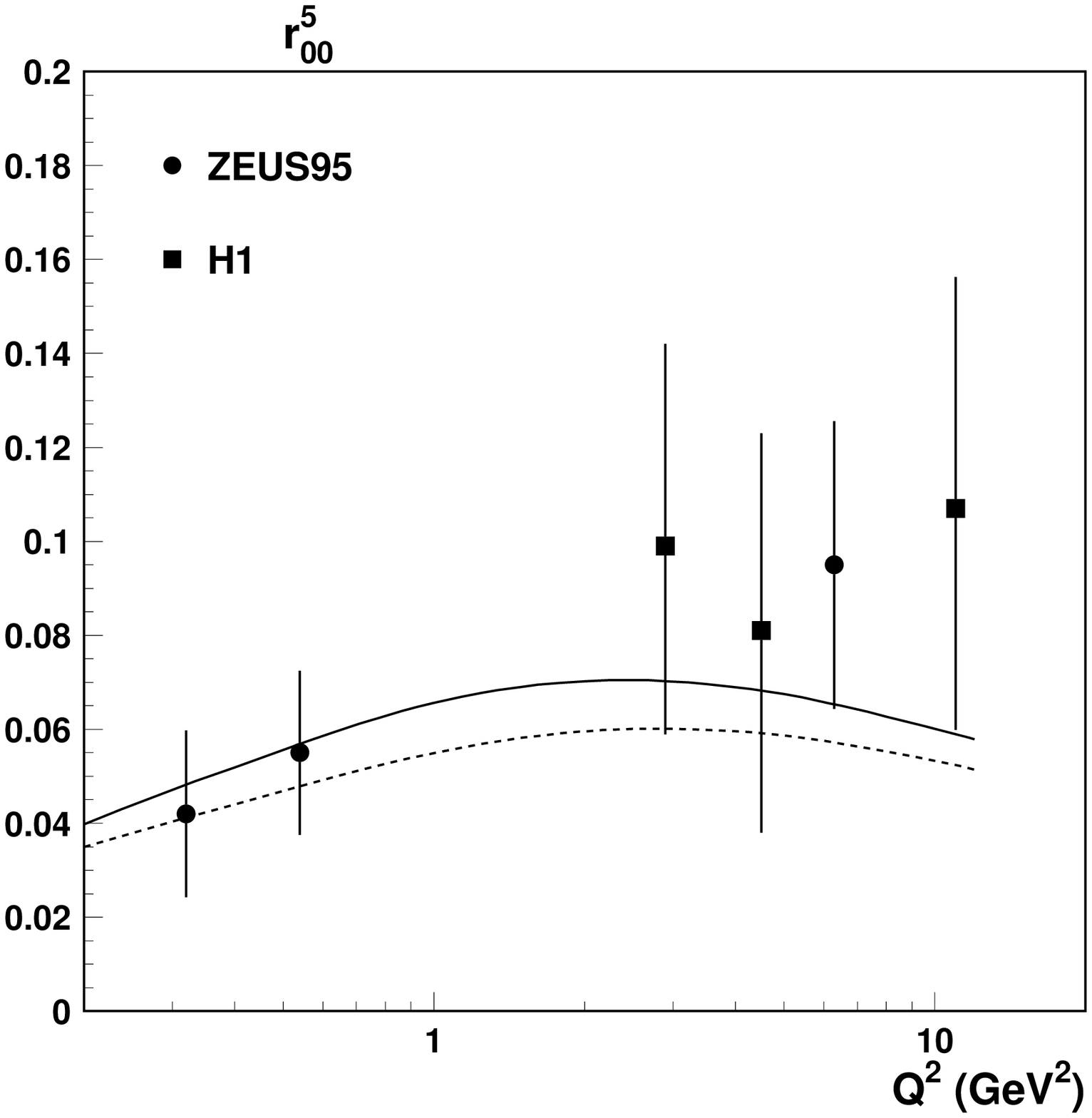}
}
\end{picture}
\end{center}
\vspace{1cm}
\caption{\protect \it
The dependence of  Born (dashed line) and
radiative corrected
(solid  line) spin-density  matrix elements  on $Q^2$ under
the   kinematic   conditions   of   H1/ZEUS  experiments.
 }
\label{fig1k}
\end{figure}
For the majority of the matrix elements vanishing in the SCHC
limit, radiative corrections turn out to be not greater then 1\%.
  However there are two of them,  ${\rm Re\;r}^{04}_{10}$ and
${\rm r}^{5}_{00}$, which RC
 appears to be
substantial (see fig. \ref{fig1k}). One can see, that
corrections  $\Delta {\rm Re\;r}^{04}_{10}$ and $\Delta {\rm
r}^{5}_{00}$ may reach $\sim $ 20\%.

The last result is
interesting from point
of view of the found SCHC violation: the
radiative
correction procedure reduces the observed effect.
\subsection{Elastic $ep$ scattering and  code MASCARAD}

Precise polarization measurements of nucleon form factors in electron
scattering is an essential component of new-generation electron
accelerators such as CEBAF \cite{1CEBAFothers}. This unprecedented
precision requires knowledge of higher-order electromagnetic effects at a
per-cent level. The purpose of this subsections is to analyze radiative
corrections in elastic electron proton scattering and present
computational techniques that could be used in experiments at Jefferson
Lab and other electron accelerator laboratories.

The observed cross section of the process
\begin{equation}\label{process}
 e(k_1) + N (p) \longrightarrow e'(k_2) + N(p_2),
\end{equation}
is described by one
non-trivial variable, which is usually chosen to be squared of
momentum transferred. There are two ways to reconstruct the variables,
when both lepton and nucleon final momenta are measured. In the first
way it will be denoted as  $Q^2_l=-(k_1-k_2)^2$, and for second one it
is  $Q^2_h=-(p_2-p)^2$. It is clear that there is no difference
between these definition at the Born level. 
However 
when the radiation of the lepton is only
considered 
at the level of RC 
the symmetry is breaking. Here we will deal with both situations.

In the first case the structure of bremsstrahlung cross section looks
like
\begin{equation}
{d\sigma \over d Q^2_l} \sim \alpha^3 \int
{d^3 k \over k_0} \sum {\cal K} {\cal F}^2(Q^2_h) {\cal A}
\label{Q2l}
\end{equation}
where $\cal K$  is a kinematic coefficient calculatable exactly in the
lowest order. It depends on photon variables. ${\cal F}^2$ is a bilinear
combination of nucleon
formfactors dependent on $Q^2_h$ only, which is function of photon
momentum. Usually only the final momenta are measured in the part of the
space. It is controlled by the function of acceptance ${\cal A}$,
which is 1 or
0 in depending whether the final particles make it in detectors or not.
The integral
(\ref{Q2l}) should not be analytically calculated for two reasons.
The first one is dependence of formfactors on $Q^2_h$. We avoid to use some
specific model for them. The second one is acceptance usually very
complicated function of kinematic variable dependent on photon
momentum.

For the second version of reconstruction of transfer momentum squared
the structure of the cross section is
\begin{equation}
{d\sigma \over d Q^2_h} \sim \alpha^3 \sum {\cal F}^2(Q^2_h)\int
{d^3 k \over k_0} {\cal K}  {\cal A}
\label{Q2h}
\end{equation}
In this case the squared formfactor is not dependent on photon momentum and
for 4$\pi$ kinematics (${\cal A}=1$) this integral can be calculated
analytically. In the experimental conditions at JLab \cite{1CEBAFothers},
both
final electron and proton were detected in order to reduce background. However
elastic scattering kinematics was restored by the final proton kinematics, while
electron momentum was integrated over. Therefore, formalism related with
Eq.(\ref{Q2h}) applies for this case.

Polarization effects are described by polarization four-vectors
of the lepton ($\xi$) and proton ($\eta$) which can be expanded
over the measured momenta $k_1$, $p_1$ and $p_2$. The polarization
vector of the lepton has been defined earlier by eq.(\ref{vecpol}) as
well as polarization vector of the proton can be expand over four
momenta of scattering particles
\begin{eqnarray}
 \eta&=&2a_{\eta}k_1+b_{\eta}q+c_{\eta}(p_1+p_2),
\end{eqnarray}
where the expansion coefficients defined by
the different polarization states of the target.

As a result the Born cross section for the process
(\ref{process}) can be written in the form
\begin{equation}
\frac{d\sigma_0}{dQ^2}={2\pi\alpha^2 \over S^2 Q^4} \sum_{i=1}^4
\theta_B^i
{\cal F}_i,
\label{si0}
\end{equation}
where
\begin{equation}
\begin{array}{ll}
\displaystyle
\theta_1^0 = 2Q^2,
&\displaystyle
\theta_3^0 = -{2Q^2\over M} (Q^2 a_{\eta}+(2S-Q^2)c_{\eta}),
\nonumber \\
\displaystyle
\theta_2^0 = {1\over M^2} (S^2-Q^2S-M^2Q^2),
&\displaystyle
\theta_4^0 = {Q^4\over M^3} (2S-Q^2) (a_{\eta}-2b_{\eta}).
\nonumber
\end{array}
\end{equation}
The structure functions ${\cal F}_i^2$ are bilinear
combinations of the nucleon
formfactors dependent on $Q^2_h$ only
\begin{equation}
\begin{array}{ll}
\displaystyle
{\cal F}_1=4 \tau_p M^2 G_M^2,
&\displaystyle
{\cal{F}}_3=-2M^2G_EG_M,
\nonumber \\
\displaystyle
{\cal F}_2=4 M^2 \frac{G_E^2+\tau_p G_M^2}{1+\tau_p},
&\displaystyle
{\cal{F}}_4=-M^2G_M{G_E-G_M\over 1+\tau_p}.
\end{array}
\end{equation}
where $\tau_p=Q^2/4M^2$.

According to \cite{pep1} the cross section that  takes into
account radiative effects within the leptonic variables can
be written as
\begin {equation}
d\sigma _{obs}
=
d\sigma _0
e^{\delta_{inf}}
(1+ \delta_{VR}+\delta_{vac})+
d\sigma _F.
\label{eq1e}
\end {equation}
Here the corrections
 $\delta_{inf}$  and   $\delta_{vac}$  come from  radiation of soft
photons and effects of vacuum polarization.
The correction  $\delta_{VR}$ is an infrared-free sum of factorized
parts of real and virtual photon radiation.
The infrared-free contribution of bremsstrahlung
process can be presented as integral over
three variables: inelasticity
$v=\Lambda^2-M^2$ ($\Lambda=p+k_1-k_2$),
$\tau=kq/kp$ and angle $\phi_k$ between planes
(${\bf q}$, ${\bf k}$) and  (${\bf k_1}$, ${\bf k_2}$)
\begin{eqnarray}
\label{sir}
d\sigma_F&=&-{\alpha^3 \over 2S^2}dQ^2_l
\int\limits_0^{v_m} {dv}
\int\limits_{\tau_{min}}^{\tau_{max}}
\frac{d\tau}{1+\tau}
\int\limits_{0}^{2\pi} {d\phi_k}
\sum_i
\biggl[\sum_{j=1}^3 {\cal A} R^{j-2}\theta_{ij} {{\cal F}_i \over Q^4_h}
 -4F^0_{IR}\theta_{i}^B {{\cal F}_i^0 \over RQ^4_l}\biggr]
.
\end{eqnarray}
Here $R=v/(1+\tau)$ and $\cal A$ is integrated over the $\phi$
acceptance function.
The quantities $\theta_{ij}$ are similar to ones used in eq.
(\ref{eq25}).
However there they are integrated over $\phi_k$. We keep this
integration because of possible dependence of acceptance function on this
angle. The explicit expressions of the $\theta$-functions in
(\ref{sir}) are discussed in Appendix of \cite{pep1}.

Since within the hadronic variables the integration over the
phase space of the real photon can be performed analytically, the
observable cross section for this situation has more simple form
\begin {equation}
d\sigma _{obs}
=\frac{\alpha ^3}{4} \frac {d Q_h^2}{S^2Q_h^4}
\sum_{i=1}^{4}[\theta _i^F+4(\delta ^{el}+\delta_{vac}^l+\delta_{vac}^h)
\theta _i^0]{\cal F}_i+d\sigma _0.
\label{eq1e2}
\end {equation}
The explicit expressions for $\theta _i^F $ and $\delta ^{el}$ can
be found in \cite{pep2}.

On the basis of formalism presented in this subsections
the FORTRAN code MASCARAD developed.
This code uses Monte-Carlo methods to
calculate RC to the observable quantities in
polarized $ep$ scattering measurements.

Here we consider RC to some obsevables within
two polarization measurements:
\begin{enumerate}
\item The initial proton is polarized and the final electron is measured
to
reconstruct $Q^2$. As a result there are four experimental situations
for asymmetry definition: the target is polarized along (perpendicular) to
beam or $\vec q$ ($q=p_2-p$). Corresponding polarization
4-vectors are denoted as
$\eta_L$ ($\eta_T$) or
$\eta_L^q$ ($\eta_T^q$).
\item Polarization and momentum of the final proton are measured. Two
polarization states should be considered: the final proton is polarized
along ($\eta'_L$) and perpendicular ($\eta'_T$) to  $\vec q$.
\end{enumerate}

It is natural that for the fist type of measurement we have to
calculate RC within the leptonic and for the second type
within hadronic variables.

\begin{figure}
\unitlength 1mm
\vspace{3cm}
\begin{picture}(80,90)
\put(5,12){
\epsfxsize=14cm
\epsfysize=14cm
\epsfbox{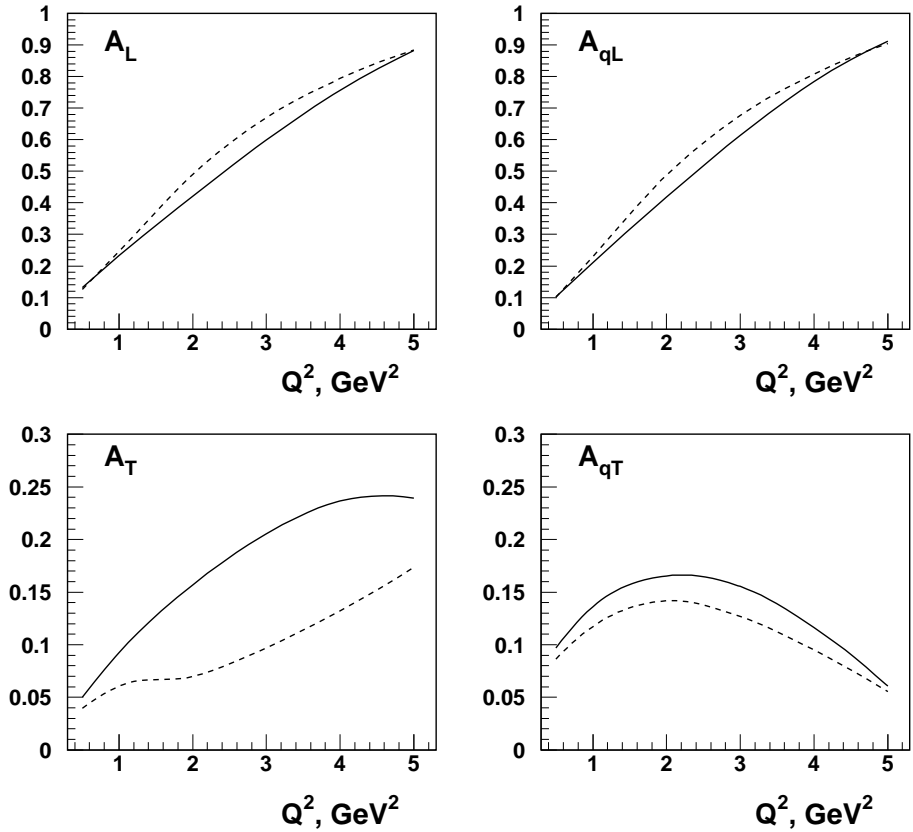}
}
\end{picture}
\vspace{-2cm}
\caption{\protect \it
Born (solid line) and observed (dashed line) asymmetries vs
$Q^2$. No kinematic cuts on inelasticity were used. $E=4$ GeV.}
\label{Fig2}
\end{figure}
\begin{figure}
\unitlength 1mm
\begin{center}
\begin{picture}(160,80)
\put(0,-10){
\epsfxsize=8cm
\epsfysize=8cm
\epsfbox{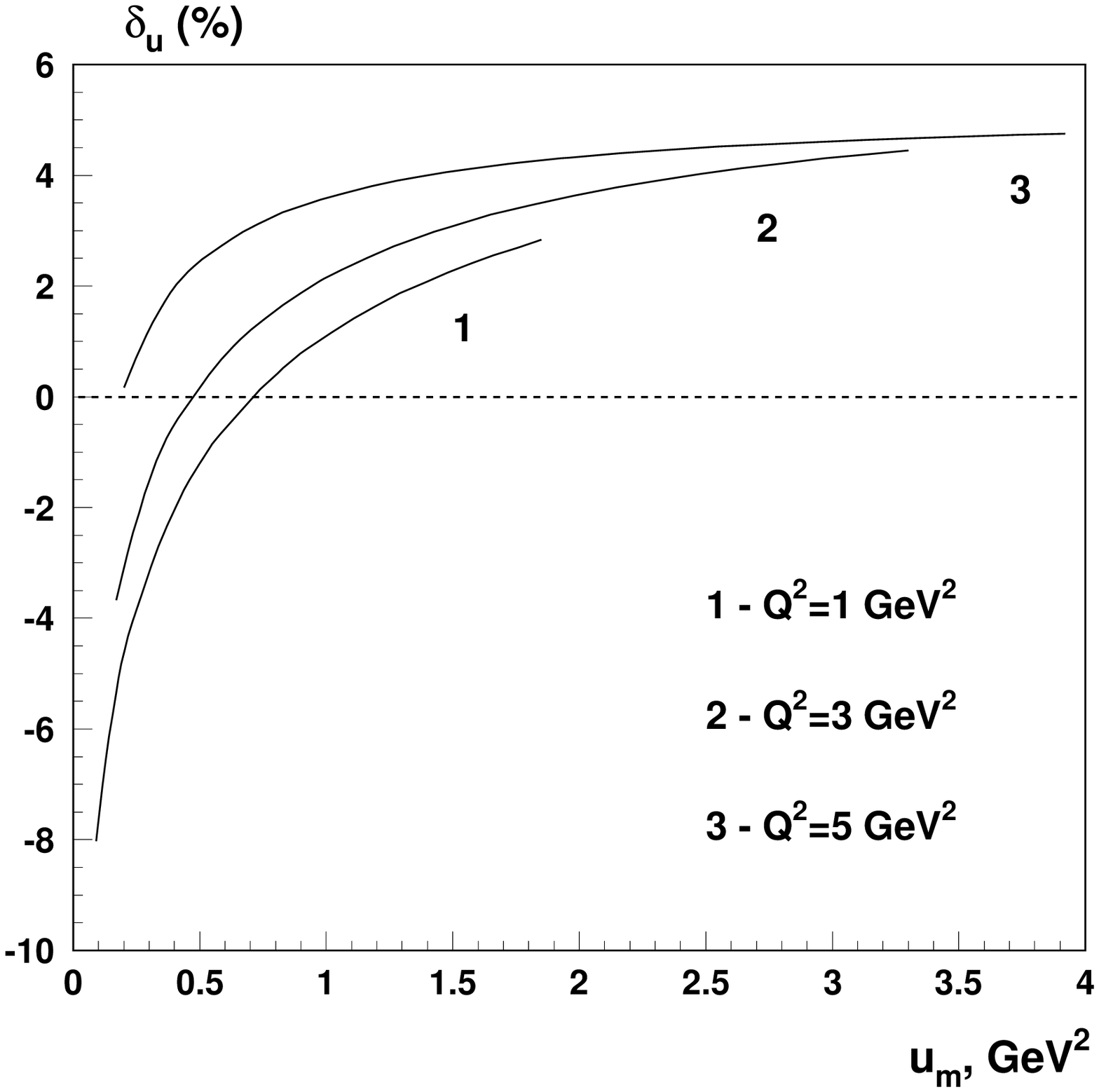}
}
\put(80,-10){
\epsfxsize=8cm
\epsfysize=8cm
\epsfbox{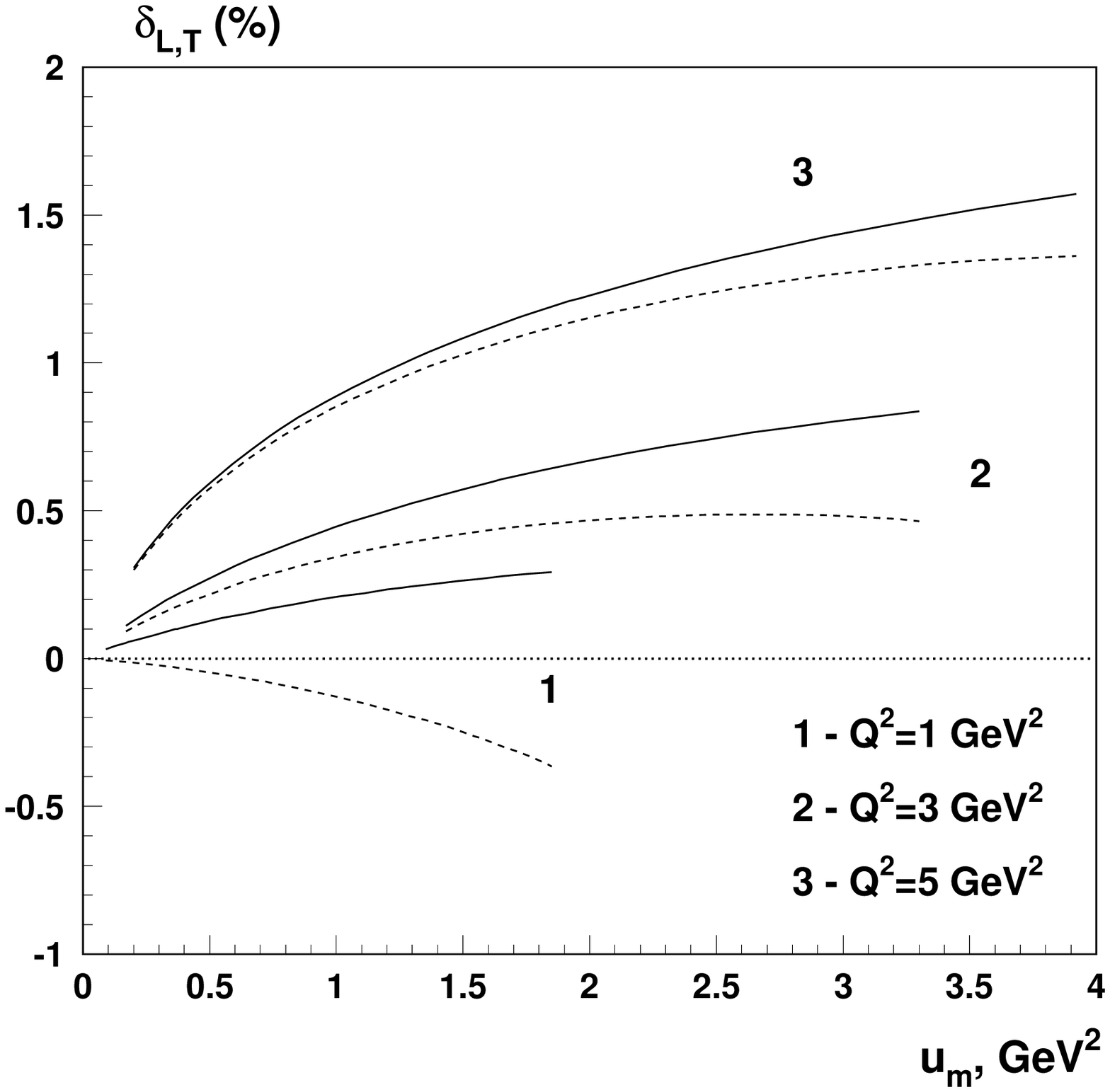}
}
\end{picture}
\end{center}
\caption{\label{fig1}
\protect \it
Radiative corrections to the unpolarized cross section (left plot) and
polarization asymmetries (right plot) defined in (\protect\ref{deltas}).
Solid and dashed  lines corresponds to
longitudinal and transverse cases. $S$=8 GeV$^2$.
}
\end{figure}

Both the spin averaged and spin dependent parts of the cross section
($\sigma^u$ and $\sigma^p$) can be
presented as
\begin{equation}\label{first}
\sigma_{obs}^{u,p}=(1+\delta)\sigma_0^{u,p}+\sigma_R^{u,p}.
\end{equation}
Both the factorized correction $\delta$ and unfactorized cross section coming
from bremsstrahlung process contribute to cross section.

Absolute and
relative
corrections to asymmetry can be defined as (see (\ref{first}))
\begin{eqnarray}\label{deldef}
\Delta A_i&=&A_i-A_{i0}={(1+\delta)\sigma_0^p+\sigma_R^p \over
(1+\delta)\sigma_0^u+\sigma_R^u} - {\sigma_0^p  \over
\sigma_0^u }\\
\Delta_i&=&{A_i-A_{i0} \over A_{i0}}=
{\delta_p-\delta_u \over 1+\delta+\delta_u}
\end{eqnarray}
where index $i$ runs over all considered cases: $i=L,T,qL,qT$;
$\delta_{u,p}=\sigma_{R}^{u,p}/\sigma_{0}^{u,p}$.  Here the correction $\delta$
is usually large because of contributions of leading logarithms. However it
exactly canceled in numerator of expression for correction to asymmetry. That
is a reason why the correction to cross section can be large while the 
correction to the
asymmetry is relatively small.

Born and observed asymmetries are presented in fig. \ref{Fig2}.
There were
no cuts  for the missing mass. As a
result
due to
unfactorizing properties
hard photon contributions gives different contributions to
spin averaged and spin dependent part of cross sections.
For the longitudinal asymmetries $\delta_p>\delta_u$ and
there are positive contributions to RC. Opposite situation with
the transverse
asymmetries.

For the second type of measurement let us define the relative
corrections to the observables in the current experiments:
\begin{eqnarray}\label{deltas1}
&&
\delta_u={\sigma_u^{obs} \over \sigma^b_u} -1, \qquad
\delta_{L,T}=\left[ \frac{\sigma_p^{obs}}{\sigma_u^{obs}}
-\frac{\sigma_p^{b}}{\sigma_u^{b}} \right]
\left[ \frac{\sigma_p^{b}} {\sigma_u^{b}} \right]^{-1}
={\sigma_u^{b} \over \sigma^{obs}_u} {\sigma_p^{obs} \over \sigma^{b}_p}-1
\end{eqnarray}

The first quantity $\delta_u$ is the relative correction to the
unpolarized part of cross section. The $\delta_{L,T}$ give contribution to
polarization asymmetries measured by rotating the polarization states of
the initial protons.
The correction to the unpolarized part of cross
section is presented in fig. \ref{fig1}a. Its behavior is quite typical.
For the very hard inelasticity cut ($u_m \ll Q^2$) the positive
contribution is suppressed due to real bremsstrahlung, so there is only
negative loop correction contributing to cross section. Different ending
values for the curves corresponds to different kinematically allowed
regions.
\section*{Conclusion}
\addcontentsline{toc}{section}{Conclusion}

Some radiative effects in
lepton scattering on polarized and unpolarized nucleons and light
nuclei have been briefly reviewed. The processes of inclusive,
semi-inclusive, diffractive and elastic scattering are considered.
The explicit expressions
for QED, electroweak and QCD lowest-order correction within
Bardin-Shumeiko approach have been presented. Besides the
next order correction in LO has been estimated.

Basing on the presented formulae some FORTRAN codes have been
created which allow one to provide the procedure of RC of
experimental data in the current and future experiments as well.
More popular of them are POLRAD \cite{P20} and RADGEN \cite{RAD}.
The version 2.0 of FORTRAN code POLRAD  provides complete RC procedure in
inclusive and semi-inclusive experimental setup.
Monte Carlo generator RADGEN creating from FORTRAN code POLRAD 2.0
provides event simulation with taking
into account radiative effects in various experimental setup on
inclusive DIS. Codes DIFFRAD and MASCARAD calculates RC in diffractive and
elastic processes.

The obtained results and FORTRAN codes constructed on their basis
are widely used in experimental collaborations at CERN, DESY, SLAC and
TJNAF.

\section*{Acknowledgement}

One of us (I. A.) thanks the US Department of Energy for
support under contract DE-AC05-84ER40150.

\end{document}